\documentclass[usenatbib]{mn2e}
\usepackage{aas_macros}

\usepackage{color}

\usepackage[dvips]{graphics}

\newif\ifAMStwofonts

\title[Submm properties of NIR-selected galaxies in SHADES/SXDF]{
The SCUBA HAlf Degree Extragalactic Survey (SHADES) -- V. Submillimetre 
properties of near-infrared--selected 
galaxies in the Subaru/{\it XMM--Newton} deep field}

\author[T. Takagi et al.]
 {T. Takagi$^{1,2,3}$\thanks{E-mail: takagi@ir.isas.jaxa.jp}, 
 A.M.J. Mortier$^{1,8}$,  K. Shimasaku$^{4}$,
 K. Coppin$^{5,6}$, A. Pope$^{5}$, 
\newauthor
 R. J. Ivison$^{7}$, H. Hanami$^{9}$, S. Serjeant$^{1}$, D.L. Clements$^{2}$, 
 R. S. Priddey$^{10}$,
\newauthor
J. S. Dunlop$^{8}$, T. Takata$^{11}$, I. Aretxaga$^{12}$, S. C. Chapman$^{13}$,
S.A. Eales$^{14}$, 
\newauthor
D. Farrah$^{15}$, 
G.L. Granato$^{16}$, M. Halpern$^{5}$,
D.H. Hughes$^{12}$, 
E. van Kampen$^{17}$,  
\newauthor 
D. Scott$^{5}$, K. Sekiguchi$^{11}$, 
I. Smail$^{6}$, M. Vaccari$^{2}$\\ 
$^1$ 
Centre for Astrophysics and Planetary Science, University of Kent,
Canterbury, Kent CT2 7NR \\ 
$^2$ 
 Blackett Laboratory, Imperial College London, 
Prince Consort Road, London, SW7 2BZ \\
$^3$ 
Institute of Space and Astronautical Science, Japan Aerospace Exploration Agency, 
Sagamihara, Kanagawa 229 8510, Japan\\
$^4$
Department of Astronomy, School of Science, University of Tokyo, 
Tokyo 113-0033, Japan \\
$^5$
Department of Physics \& Astronomy, University of British Columbia, 6224 
Agricultual Road, Vancouver, B.C., V6T 1Z1, Canada\\
$^{6}$
Institute for Computational Cosmology, Durham University, South Rd, 
Durham, DH1 3LE \\
$^7$ UK ATC, Royal Observatory, Blackford Hill, Edinburgh EH9 3HJ \\
$^8$ SUPA\footnote{Scottish Universities Physics Alliance},
Institute for Astronomy, University of Edinburgh, Royal Observatory, 
Edinburgh EH9 3HJ \\
$^9$
Physics Section, Faculty of Humanities and Social Sciences, Iwate
University, Morioka, 020-8550, Japan \\
$^10$
Department of Physics, Astronomy \& Mathematics, 
University of Hertfordshire, College Lane, Hatfield, Hertfordshire AL10 9AB \\
$^{11}$
Subaru Telescope, National Astronomical Observatory of Japan, 650 N.\ A'ohoku 
Place, Hilo, HI 96720, USA \\
$^{12}$
Instituto Nacional de Astrof\'isica, \'Optica y Electr\'onica (INAOE), 
Apartado Postal 51 y 216, 72000 Puebla, Pue., Mexico\\
$^{13}$ 
California Institute of Technology, 1200 East California Boulevard,
Pasadena, CA 91125 \\
$^{14}$
Cardiff School of Physics and Astronomy, Cardiff University, 5, 
The Parade, Cardiff, CF24 3YB\\
$^{15}$
Department of Astronomy, Cornell University, Space Sciences Building, 
Ithaca, NY 14853, USA\\
$^{16}$
Istituto Nazionale di Astrofisica, Osservatorio Astronomico di Padova, 
Vicolo dell'Osservatorio 5, I-35100 Padova, Italy \\
$^{17}$
Institute of Astro- and Particle Physics, 
University of Innsbruck, Technikerstr. 25,
A-6020, Innsbruck, Austria \\
}

\date{}

\begin{document}

\maketitle


\begin{abstract}
We have studied  the submillimetre (submm) properties of the following 
classes of near-infrared\,(NIR)-selected massive galaxies at high redshifts:
$BzK$-selected star-forming galaxies ($BzK$s); distant red galaxies 
(DRGs); and extremely red objects (EROs). 
We used the SCUBA HAlf Degree Extragalactic Survey (SHADES), 
the largest uniform submm survey to date. 
Partial overlap of SIRIUS/NIR images and SHADES in SXDF has allowed us 
to identify 4 submm-bright NIR-selected galaxies, which are detected 
in the mid-infrared, 24\,$\mu$m, and the radio, 1.4\,GHz. 
We find that all of our submm-bright NIR-selected galaxies 
satisfy the $BzK$ selection criteria, i.e. $BzK \equiv 
(z-K)_\mathrm{AB}-(B-z)_\mathrm{AB} \ge -0.2$, except for 
one galaxy whose $B-z$ and $z-K$ colours are however close to the $BzK$
colour boundary. 
Two of the submm-bright NIR-selected galaxies 
satisfy all of the selection criteria we considered, i.e. 
they belong to the $BzK$-DRG-ERO overlapping population, 
or `extremely red' $BzK$s. 
Although these extremely red $BzK$s are rare (0.25\,arcmin$^{-2}$), 
up to 20\,\% of this population could be submm galaxies. This fraction is 
significantly higher than that found for other galaxy populations studied here.
Via a stacking analysis, we have detected the 850-$\mu$m flux of 
submm-faint $BzK$s and EROs in our SCUBA maps. 
While the contribution of $z\sim 2$ $BzK$s 
to the submm background is 
about 10--15\,\% and similar to that from EROs typically at $z\sim 1$, 
$BzK$s have a higher fraction ($\sim 30$\,\%) 
of submm flux in resolved sources compared with 
EROs and submm sources as a whole. 
From the SED fitting analysis for both submm-bright and submm-faint $BzK$s, 
we found no clear signature that submm-bright $BzK$s are experiencing a 
specifically luminous evolutionary phase, compared with submm-faint $BzK$s. 
An alternative explanation might be that submm-bright $BzK$s are more 
massive than submm-faint ones. 
\end{abstract}


\section{Introduction}\label{sec:introduction}

The first 3--4 billion years in the history of the universe 
appears to have been a very important epoch for the formation of present-day
massive galaxies, in particular for 
elliptical galaxies and bulges with masses 
of  $M\ga 10^{11} M_\odot$. Recent wide field near-infrared (NIR) 
surveys in conjunction with multi-band optical images or spectroscopy
have resulted in a profusion  of samples of 
massive galaxies at high redshifts 
\citep[e.g.][]{
2003ApJ...587L..79F,
2003ApJ...594L...9F,
2003ApJ...587...25D,
2004A&A...424...23F,
2004Natur.430..184C,
2004Natur.430..181G,
2004ApJ...600L.127D,
2005ApJ...619L.135J,
2005ApJ...624L..81L,
2005ApJ...619L.131D,
2005ApJ...633..748R,
2006ApJ...638...72K,
2006ApJ...640...92P,
2007arXiv0705.2831D,
2007arXiv0705.2438A,
2007arXiv0704.1600P}.
At $z\sim 2$, the physical properties of massive galaxies are
very different from those of present-day ones, i.e. a large fraction of them  
are intensively star-forming galaxies \citep[e.g.][]{2005ApJ...631L..13D}, 
rather than passively evolving, like nearby ones. 
A notable characteristic of these massive star-forming galaxies is
that, in some cases, a large total stellar mass $\sim 10^{11} M_\odot$ 
could be formed in the time-scale 
of a single starburst event, $\sim 0.1$ Gyr \citep{2005ApJ...630...82P,
2006ApJ...637..727C,2006ApJ...640...92P}. 

Massive, intensively star-forming galaxies have also been found 
in submillimetre (submm) surveys \citep{1997ApJ...490L...5S,
1998Natur.394..241H,1999ApJ...515..518E,2002MNRAS.331..817S,
2003MNRAS.344..385B,2005MNRAS.363..563M,2006MNRAS.372.1621C}. 
These submm galaxies turn out to be typically starbursts or starburst/AGN 
composite systems at $z\sim 2$ 
\citep[][and references therein]{2005ApJ...622..772C,2005ApJ...632..736A}, 
which are massive \citep{2004ApJ...617...64S,2005MNRAS.359.1165G}. 
They are likely to become passively evolving massive galaxies 
at lower redshifts \citep[e.g.][]{2004ApJ...616...71S,2004ApJ...611..725B,
2004MNRAS.355..424T}. The star formation rates (SFRs) of 
submm-bright galaxies are estimated to be 
extraordinarily high, $\sim 10^3$ M$_\odot$ yr$^{-1}$ 
\citep[e.g.][]{2005MNRAS.359.1165G,2005ApJ...632..736A,2005ApJ...622..772C,
2006MNRAS.370.1185P,2006ApJ...650..592K}, 
enough to produce a massive elliptical galaxy ($>3$ L$^*$) within 
$\sim 1$ Gyr. 
Determining the nature of these star-forming galaxies at high redshifts 
may be the key to understanding the process of massive galaxy formation. 

Efficient methods for selecting massive star-forming 
galaxies at high redshifts 
with one or two optical/NIR colours have played a significant role 
in studying the properties of massive galaxy populations. 
These include $BzK$-selected galaxies (selected in $B-z$ and $z-K$ colours), 
distant red galaxies (selected with $J-K$ colour), and 
extremely red objects (selected with $R-K$ or $I-K$ colour). 
Hereafter we call these galaxy populations NIR-selected galaxies or NIRGs. 

\cite{2004ApJ...600L.127D} proposed an effective method 
for selecting massive star-forming galaxies at $1.4 \la z \la 2.5$ 
along with the selection of 
passively evolving galaxies at a similar redshift range
in a colour-colour diagram of $B-z$ vs $z-K$. 
It is expected to be reddening independent, i.e. applicable to heavily 
obscured galaxies like ultraluminous infrared galaxies (ULIRGs). 
Their luminosities at UV (reddening corrected),
infrared, X-ray and radio indicate that $BzK$-selected star-forming 
galaxies ($BzK$s)\footnote{Since 
we are studying cross-identifications between $BzK$-selected galaxies 
and submm galaxies, star-forming galaxies selected with this method 
are simply referred to as $BzK$-selected galaxies or $BzK$s.} 
with $K<20$ are typically ULIRGs with 
SFRs of $\sim 200$\,M$_\odot$\,yr$^{-1}$ and reddening of 
$E(B-V)\sim 0.4$ \citep{2004ApJ...617..746D,2005ApJ...631L..13D,
2005ApJ...633..748R,2006ApJ...638...72K}. 

In an alternative approach, \cite{2003ApJ...587L..79F} employed a colour cut of 
$(J-K_\mathrm{s})_\mathrm{Vega}>2.3$ 
to select distant red galaxies (DRGs). This colour criterion selects both 
passively evolving and heavily obscured star-forming galaxies at $2<z<4.5$ with 
strong Balmer or 4000\AA\ breaks. 
Recent {\it Spitzer} imaging at 24\,$\mu$m suggests that 
the bulk of DRGs are dusty star-forming galaxies 
\citep{2006ApJ...636L..17W, 2006ApJ...640...92P}. 
Similar to $BzK$-selected 
star-forming galaxies, DRGs are likely to be ULIRGs at $z\sim 2$ 
\citep[e.g.][]{2005ApJ...632L...9K}, but to also include a 
significant fraction of less luminous dusty star-forming galaxies at $1<z<2$ 
\citep{2006A&A...453..507G}.
Submm galaxies, another sample of ULIRGs at $z\sim 2$, 
could have a large overlap with $BzK$s or DRGs, 
but with systematically higher 
SFR than the average \citep[e.g.][]{2006ApJ...637L...5D}. 

Extremely red objects (EROs) defined with $R-K$ or 
$I-K$ colours have received particular attention as possible optical-NIR counterparts 
of submm galaxies \citep[e.g.][]{1999MNRAS.308.1061S,2003ApJ...597..680W,
2005MNRAS.358..149P}, since submm galaxies are expected to be heavily 
obscured by dust and therefore red. However, it turns out that only 
10--30\,\% of submm galaxies, whose optical counterparts 
are identified with associated radio sources, are EROs 
\citep{2002MNRAS.337....1I,2003ApJ...597..680W,2004MNRAS.355..485B,
2004ApJ...616...71S}. The median redshift of EROs with 
$K_\mathrm{Vega} < 19.2$ is $z=1.1\pm0.2$ \citep{2002A&A...381L..68C}, 
and therefore lower than that 
of resolved submm-bright galaxies, $BzK$s and DRGs. Nevertheless, there is 
certainly some overlap between their redshift distributions.

It is important to understand the the physical relationships between 
these high-$z$ massive star-forming galaxies which are selected in different 
ways \citep[e.g. see][]{2005ApJ...633..748R}. 
The surface density of $BzK$s, DRGs and EROs is similar to each other, about 1--2 arcmin$^{-2}$ with 
$K_\mathrm{Vega} \la 20$ \citep{2004ApJ...617..746D,2002A&A...381L..68C,2003ApJ...587L..83V}.
A similar surface density has been found for submm galaxies with 
$S_{850\mu \mathrm{m}} \ga 1$\,mJy \citep[e.g.][]{1999ApJ...512L..87B}, 
although the surface density of bright ($S_{850\mu \mathrm{m}} \ga 8$\,mJy) 
submm galaxies in `blank sky' surveys 
is at least an order of magnitude smaller. 
Understanding the physical origin of the overlap 
between submm galaxies and NIR-selected 
galaxies is not straightforward, given the diverse optical-NIR properties of 
submm galaxies, which include those similar to 
less-attenuated UV-selected galaxies 
as well as EROs \citep[e.g.][]{2002MNRAS.331..495S,2004ApJ...616...71S}. 

Here we study the submm properties of $BzK$s, DRGs and EROs 
by using the SCUBA HAlf Degree Extragalactic Survey 
\citep[SHADES --][]{2005MNRAS.363..563M}, providing a larger sample 
of submm galaxies for investigating the 
relation between submm galaxies and NIRGs than the study of 
\cite{2005ApJ...633..748R} for galaxies in the GOODS-North field. 
SHADES has mapped two 
regions 
in the Lockman Hole and in the Subaru/{\it XMM--Newton} deep field (SXDF). 
In this study, we focus on a 93$\,$arcmin$^2$ sub-region of 
the SXDF field of SHADES, in which 
both optical ($BVRi'z'$) and NIR ($JHK_\mathrm{s}$) photometry are
already available. 
We describe the data in Section 2 and 
the sample of NIR-selected galaxies in Section 3. 
In Section 4, we identify submm-bright NIR-selected galaxies and 
assess the overlap between submm galaxies and NIR-selected galaxies. 
In Section 5, we present the results of a statistical detection of 
NIR-selected galaxies in the SCUBA map using a stacking analysis. 
In Section 6, we analyse the spectral energy distributions (SEDs) of submm-bright NIR-selected 
galaxies and also an average SED of submm-faint NIRGs, in order 
to investigate their physical properties. 
Summary of our study is given in Section 7. 
Throughout this paper, we adopt the cosmology of 
$\Omega_\mathrm{m} =0.3$, $\Omega_\Lambda =0.7$ and 
$H_0 =70$$\,$km$\,$sec$^{-1}$$\,$Mpc$^{-1}$.
All the magnitudes are given in the AB system, unless 
otherwise noted.  

\section{The data}

\subsection{Optical/NIR data in the SXDF}
We have used the optical and NIR imaging data of the SXDF 
centred at [$2^\mathrm{h}18^\mathrm{m}00^\mathrm{s}$, 
$-5^\circ 00'00^{\prime\prime}$ (J2000)] 
published by \cite{2003PASJ...55.1079M}. The NIR data were 
obtained with the University of Hawaii 2.2\,m telescope with 
the Simultaneous 3-colour InfraRed Imager for Unbiased Survey 
(SIRIUS; \citealt{2003SPIE.4841..459N}). 
The final image covers 
an area of 114$\,$arcmin$^2$ in the $H$- and $K_\mathrm{s}$-bands. 
For the $J$-band, we obtained data only for 77$\,$arcmin$^2$, 
since a quarter of the field of view was not operational. 
The 5\,$\sigma$ detection limits of the final NIR image 
are $J=22.8$, $H=22.5$ and $K_\mathrm{s} = 22.1$ mag, through a 
$2^{\prime\prime}$ diameter aperture. 
From the final image in the $K_\mathrm{s}$-band, we have detected 1308 
objects having $K_\mathrm{s} < 22.1$ mag 
using SExtractor \citep{1996A&AS..117..393B}. 
Throughout this work we use the total magnitudes in the optical-NIR bands 
measured with the MAG\_AUTO algorithm in SExtractor. 

For these $K_\mathrm{s}$-band detected objects, we made a multi-band catalogue 
using optical images of Subaru/Suprime-cam \citep{2002PASJ...54..833M} 
in $B$, $V$, $R$, $i'$, $z'$ bands\footnote{Our data were obtained during the 
commissioning phase of the Suprime-cam.}.
The limiting magnitudes (5\,$\sigma$) of our images are 
$B = 27.1$, $V=25.9$, $R=26.3$, $i'=25.7$, $z'=25.0$ mag 
through a $2^{\prime\prime}$ diameter aperture. 
The \rm{FWHM} of the point-spread function is $0.98^{\prime\prime}$ in both 
the 
optical and NIR images. 
Stars were identified with the colour 
criterion of $B-K_\mathrm{s} < 1.583 (B-i') - 0.5$ and 
FWHM $\le 1.2^{\prime\prime}$ \citep{2003PASJ...55.1079M}, and 
removed from the analysis.

\subsection{Submm/radio data from SHADES}
Details of the survey design and observing strategy with JCMT/SCUBA are 
given in \cite{2005MNRAS.363..563M}. Here we describe the SHADES survey 
only briefly.  
Using the SCUBA instrument on the JCMT we have obtained jiggle maps
of the SXDF and also the Lockman Hole in grade 2--3 weather ($\tau_\mathrm{CSO} = 0.05 - 0.1$). 
The \rm{FWHM} of the SCUBA beam is 14.7$''$ and 7.8\,$^{\prime\prime}$ at 
850 and 450\,$\mu$m, respectively.
The SHADES observations with SCUBA were continued from
December 2002 to the decommissioning of 
SCUBA in June 2005. Here we use the SCUBA data acquired until 
1st Feb, 2004. 

The SCUBA data have been reduced by four independent groups in the 
SHADES consortium. Practical methods of data reduction, such as 
flux calibration, extinction correction, and map making 
depends on the reduction group and are summarised in 
\cite{2006MNRAS.372.1621C}. 
The extracted sources from four different SCUBA maps have been 
compared and then combined to produce the SHADES source catalogue, 
which is expected to be the most reliable source catalogue from 
SCUBA surveys. In our analyses, we focus on two particular SCUBA maps 
out of four, along with the SHADES source catalogue, in order to 
make the sample of submm-bright NIRGs as complete as possible. 

We have chosen to use only regions with noise less than 3\,mJy at 850\,$\mu$m.
The total area of this region is $\sim$250\,arcmin$^2$. 
The field centre for the SHADES map of the SXDF,
$02^{\rm h}$$17^{\rm m}$$57^{\rm s}.5$,
$-05^\circ$$00^{\prime}$$18^{\prime\prime}.5$ (J2000), 
is offset from that of the
SIRIUS observations and hence 
our sub-mm maps do not cover the whole region covered by the NIR data. 
In one of the four SCUBA maps (`Reduction B' in Coppin et al.),  
this overlap is 93\,arcmin$^2$, with median noise levels of 
2.0\,mJy and 18.4\,mJy at 850 and 450\,$\mu$m, respectively. 

Wide-field 1.4-GHz radio images of SXDF were obtained using the VLA during
2003. Around 60\,hr of integration were salvaged, following a prolonged
failure of the correlator, comprising 
data from the A, B and C configurations, with
an approximate 9:3:1 ratio of recorded visibilities, evenly distributed in
three pointings separated by 15\,arcmin. The final images were mosaiced
together, after correcting for the response of the primary beam. The
resulting noise level is around 7\,$\mu$Jy\,beam$^{-1}$ in the best
regions of the map (rather higher near bright, complex radio emitters) with
a synthesised beam measuring 1.87\,$\times$\,1.65\,arcsec (FWHM),
major axis 22$^{\circ}$ east of north. The data and their reduction are
described in detail in \cite{2007astro.ph..2544I}. 

\subsection{{\it Spitzer} data from SWIRE}
The whole area of our $K_\mathrm{s}$-band image is covered by the {\it Spitzer} Wide-area 
InfraRed Extragalactic survey \cite[SWIRE;][]{2003PASP..115..897L}.
We used the IRAC+24\,$\mu$m catalogue from version 2.0 data products 
(released in Summer 2005), 
available from the NASA/IPAC Infrared Service Archive (IRSA). 
We adopted the Kron fluxes, which are MAG\_AUTO fluxes from SExtractor, 
in order to compare with the ground-based optical-NIR photometry. 
This catalogue includes IRAC sources which are detected at both 
3.6\,$\mu$m ($>10\,\sigma$) and 4.5\,$\mu$m ($>5\,\sigma$). 
The astrometric error of the IRAC sources is small enough to easily 
identify counterparts of NIRGs with the angular separation of $\la 0.5''$. 
In the SIRIUS/$K_\mathrm{s}$-band region, the faintest sources at 24$\,\mu$m in 
the catalogue have a flux of $\sim$300\,$\mu$Jy. 
Cross-correlation between the SHADES sources and the SWIRE sources 
in SXDF is given in Clements et al. (2007, in preparation).

\section{Sample of NIR-selected galaxies at high redshifts} 

\subsection{EROs and DRGs} 
\begin{figure*}
  \resizebox{7cm}{!}{\includegraphics{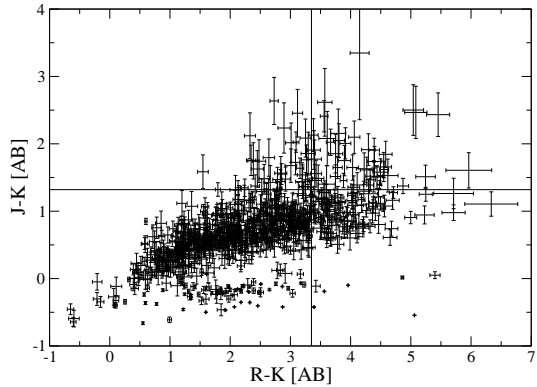}}
 \caption{$J-K_\mathrm{s}$ versus $R-K_\mathrm{s}$ for $K_\mathrm{s}$-band detected objects 
in the SIRIUS $J$-band area. The selection criteria for EROs and 
DRGs are indicated with a vertical line and a horizontal line, respectively. 
}
 \label{scat1}
\end{figure*}

Following \cite{2003PASJ...55.1079M}, we define EROs 
as objects with $R-K_\mathrm{s} > 3.35$, which is equivalent to 
$(R-K)_\mathrm{Vega} \ga 5$, i.e. the widely used criterion of EROs. 
In order to obtain the DRG sample, we use 
the threshold of $J-K_\mathrm{s}>1.32$ (i.e. $\ga 2.3$ in Vega magnitudes). 
Within the SHADES area, we have detected 201 EROs and 67 
DRGs in total, among which 39 objects satisfy 
both $R-K_\mathrm{s} > 3.35$ and $J-K_\mathrm{s} > 1.32$. 
In Figure \ref{scat1}, we show a colour-colour plot with 
$R-K_\mathrm{s}$ and $J-K_\mathrm{s}$ for $K_\mathrm{s}$-band detected objects, along 
with the adopted colour criteria. The statistics of NIRGs are summarised 
in Table \ref{tab_stat}. 

We derived a surface density of $2.18\pm 0.14$ and 
$1.09 \pm 0.12$$\,$arcmin$^{-2}$ (Poissonian errors)
for EROs and DRGs, respectively. 
By using the same data, 
\cite{2003PASJ...55.1079M} found a good agreement in 
the surface density of EROs with those of the other 
surveys with areas of $>50$$\,$arcmin$^2$ 
\citep{2002A&A...381L..68C,2000A&A...361..535D,
1999ApJ...523..100T}. 
The surface density of DRGs is 
coincidentally very similar to that derived by 
\cite{2003ApJ...587L..83V}, although the $K_\mathrm{s}$-band limiting 
magnitude in \cite{2003ApJ...587L..83V} is $\sim 0.7$ mag 
deeper than ours. Thus, we find a higher surface density of $K_\mathrm{s} \la 20$
DRGs than in \cite{2003ApJ...587L..83V}. This could be partly 
because of the contamination of the sample at $K_\mathrm{s} > 21.5$ 
where the detection of DRGs at $J$-band is possible 
only below $4$\,$\sigma$.

\subsection{BzK-selected star-forming galaxies}
\cite{2004ApJ...617..746D} proposed a joint selection of star-forming 
galaxies and passively evolving galaxies in the redshift range of 
$z=$ 1.4--2.5 in the $z-K$ vs $B-z$ diagram (hereafter `$BzK$ diagram'). 
Following \cite{2004ApJ...617..746D}, we choose 
$BzK$-selected galaxies with $BzK \ge -0.2$, where 
 $BzK \equiv (z-K) - (B-z)$. In order to adjust the 
selection in our photometric bands to that of Daddi et al., 
we compared the stellar sequence in the $BzK$ diagram with 
that of \cite{2004ApJ...617..746D}. To match 
the stellar sequence, we applied the following colour corrections: 
($B-z$)$_\mathrm{Daddi}$ = $(B-z')+0.2$ and 
$(z-K)_\mathrm{Daddi}$=$(z'-K)-0.2$. 
After this correction, we obtained 132 $BzK$-selected galaxies 
within the SHADES area. 

\begin{figure*}
  \resizebox{7cm}{!}{\includegraphics{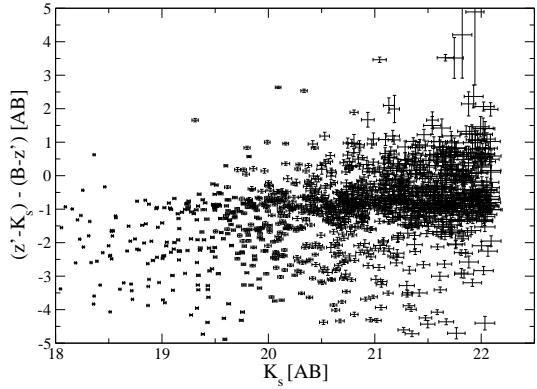}}
 \caption{$BzK$ colour vs. $K_\mathrm{s}$ magnitude for $K_\mathrm{s}$-band detected 
objects. Here the $z'-K_\mathrm{s}$ and $B-z'$ colours have {\it not} been 
colour-corrected 
in order to show the original photometric errors. The $BzK$ selection 
criterion proposed by Daddi et al. (2004) corresponds to 
$(z'-K_\mathrm{s})-(B-z')=0.2$ for the adopted colour correction (see text). 
}
 \label{scat2}
\end{figure*}

We obtained a surface density of 1.1$\,$arcmin$^{-2}$ 
for $K_\mathrm{s} < 20$ (Vega). This surface density is consistent with that of 
\cite{2005ApJ...631L..13D}, who found 169 $BzK$s within 154 arcmin$^2$, 
including X-ray detected objects, in the GOODS North region.  
Thus, we assume there are no systematic differences between our sample 
and that of \cite{2005ApJ...631L..13D}. 
In Figure \ref{scat2}, we plot the raw $BzK$ colour (i.e. with no
colour corrections) of $K_\mathrm{s}$-band detected objects 
against $K_\mathrm{s}$ magnitudes along with the photometric 
errors. 


\section{Identification of submm-bright NIRG{\sevensize s}}

\subsection{Method of identification} 
In order not to miss any candidate submm-bright NIRGs, we utilized two SCUBA 
850-$\mu$m maps which have 
different pixel size produced by independent reductions and reported in 
Coppin et al. (2006 -- Reductions B and D). 
The pixel sizes of the SCUBA 850-$\mu$m map from Reductions B and D are 
$1''$ and $3''$, respectively. One benefit 
of using SCUBA maps, rather than the SHADES source catalogue, is the possibility of 
identifying additional submm sources which are just below the limit of the SHADES 
catalogue. This is a reasonable approach, since we know the sky positions of 
targets beforehand (and see Appendix A for a statistical discussion).

We adopted the threshold of $S/N > 3$ for the detection of submm fluxes
at the positions of NIRGs. 
In order to confirm the detection of NIRGs in SCUBA maps, we need to exclude 
the possibility of; 1) chance association with a nearby SCUBA source; and 2) false positive 
detections. 
Sources in the SHADES catalogue should be quite reliable, since they are extracted 
using 4 independent reductions. 
If there are no SHADES catalogue sources corresponding to detected NIRGs, 
we need to pay special attention to check the reliability of the detection in the SCUBA 
map. To overcome these problems, we use the {\it Spitzer} images at 24\,$\mu$m and 
the VLA radio images at 1.4\,GHz as they have proved to be useful in previous 
studies \citep[e.g.][]{2002MNRAS.337....1I,2004ApJS..154..130E}.
The identification of SHADES sources has been undertaken by  
Ivison et al. (2007). We confirmed that our results are consistent with 
their identifications. For non-SHADES sources, we 
require detection at 24\,$\mu$m or in the radio for secure 
identification of submm-bright NIRGs. 

We also obtained 450-$\mu$m images simultaneously with the 850-$\mu$m images, 
although the weather conditions 
allocated for SHADES were not good enough to detect typical 850-$\mu$m 
sources at 450$\,\mu$m in the same integration time. 
We have however used the 450$\,\mu$m map to constrain the 
450-$\mu$m flux for the detected 850-$\mu$m sources. 
The reduction of 450\,$\mu$m data is fully 
described in \cite{2006MNRAS.372.1621C}. 


\subsection{Results}
Among 307 NIRGs in the SHADES/SIRIUS $K_\mathrm{s}$-band area (245 fall in $J$-band area),  
we detected 5 NIRGs at 850$\mu m$
as a result of photometry at the position of NIRGs 
in the SHADES B-map (i.e. the SCUBA 850\,$\mu$m map from Reduction B). 
We also detected 5 NIRGs at 850$\mu m$ 
in the SHADES D-map, four of which are common with 
those 
detected in the SHADES B-map. The results of the sub-mm 
detections from both SHADES maps 
are summarised in Table \ref{tab_obs1}. All of the submm-detected 
NIRGs except fo ID1390 have corresponding SHADES catalogue sources. 
Although a submm source for ID1390 
(SXDF850.62) was included in a preliminary SHADES source catalogue, 
this source did not satisfy the final significance limit adopted for the 
SHADES catalogue presented by Coppin et al. (2006). 
However, we found that ID1390 is detected both 
at 24\,$\mu$m and in the radio, and therefore we conclude that this is a 
real submm galaxy (see also Appendix A). In the 450-$\mu$m map, 
we find no convincing detections at the positions of NIRGs. This is not 
unexpected given the high noise level in the 450$\mu m$ maps. 

Three (ID300, 445, 912) out of six submm-detected NIRGs are identified 
as optical counterparts of corresponding SHADES sources in Ivison et al. (2007).
Additionally, we find ID1390 as an optical-NIR counterpart of SXDF850.62 (a 
non-SHADES source), since this galaxy is detected at 24\,$\mu$m and in the radio 
as noted above. In summary, we identify four submm-bright NIRGs. 
Figure \ref{submm} shows the 850-$\mu$m contours plotted over $K_\mathrm{s}$-band
images of submm-bright NIRGs. 
Figure \ref{submm} also shows the radio contours over $R$-band images 
of the same NIRGs. 
The remaining two submm-detected NIRGs (ID718 and ID1133) are detected neither 
in the radio nor at 24\,$\mu$m. 
Such chance associations are expected 
and the number is consistent with the estimate given in Appendix \ref{appendix}. 
For a submm source associated with ID1133, Ivison et al. (2007) found another 
reliable optical counterpart with a radio detection. 
Considering these results, we exclude both ID718 and 1133 from further analyses. 

In order to check additional candidates for optical-NIR counterparts 
of the SHADES catalogue sources, we also searched for NIR-selected 
galaxies within $7''$ radius of the SHADES positions. Although we found 
two additional associations (ID475 and 579), they are 
not detected either at 24\,$\mu$m or in the radio. Therefore, this alternative 
method does not produce any further submm-bright NIRGs for our sample. 

\subsection{Relation between submm galaxies and NIRGs}
In Table \ref{tab_obs2}, 
we tabulate the optical-NIR properties of submm-bright 
NIRGs. Note that all of the submm-bright NIRGs 
are $BzK$-selected galaxies, except for ID1390 with $BzK=-0.36$, 
although it satisfies $BzK \ge -0.2$ before the colour 
correction on $B-z$ and $z-K$. 
This also means that no EROs and DRGs which are 
clearly {\it non}-$BzK$ galaxies are detected at 850\,$\mu$m. 
The distribution of submm-bright NIRGs in 
the $B-z$ vs $z-K$ colour-colour diagram is discussed 
in detail in Section \ref{sec:bzk}. 

Two galaxies, ID300 and ID445, satisfy all the colour selection 
criteria we adopted, i.e. $BzK \ge -0.2$, $R-K > 3.35$ and 
$J-K > 1.32$. 
Hereafter, we refer to this $BzK$-DRG-ERO overlapping population as 
extremely red $BzK$-selected galaxies. 
We found only 16 extremely red $BzK$s in 
the combined $K_\mathrm{s}$ and SHADES areas, corresponding to a
surface density of only $0.25\pm 0.05$ arcmin$^{-2}$. 
This means that $12\pm 8$\,\% (2/16)\footnote{The errors on fractions, $f$, 
are given by $\sqrt{\frac{f (1-f)}{n} }$ where $n$ is the 
total number of the sample. This gives a rough estimate based on the normal 
approximation to the binomial distribution.} of extremely red $BzK$s 
are submm galaxies. Although the sample is very small, nevertheless it may
indicate that this rare NIR galaxy population has a much 
higher fraction of 
submm galaxies, compared with the other optical-NIR selected 
galaxy populations already studied. For example, the fractions 
of submm galaxies which are $BzK$s, EROs and DRGs in our sample are  
4/132 ($3\pm 2$ \%), 3/201 ($1.5\pm 1$ \%) and 
3/67 ($5\pm 3$ \%), respectively. 

We found that there are 20 (13) SHADES sources in the 
SIRIUS/$K_\mathrm{s}$-band ($J$-band) area, among which 
three are $BzK$-selected 
and two satisfy both the ERO and DRG selection. The resulting 
fractions of $BzK$s, EROs and DRGs in our sub-sample of 
SHADES sources are $15\pm 8$~\%, 
$10\pm 7$ \% and $15\pm 10$ \%, respectively. 
Thirteen SHADES sources out of 20 have a robust radio 
identification (Ivison et al. 2007). 
If we confine our attention to SHADES sources with a robust radio 
identification, the above fractions of NIRGs are 
higher by a factor of $\sim$1.5. 

\cite{2002MNRAS.337....1I} found an ERO 
fraction of 33 \% (6/18) in SCUBA sources with robust radio 
identification from the 8-mJy survey, in which one ERO is 
too faint to be detected at our detection 
limit in the $K_\mathrm{s}$-band. 
\cite{2005ApJ...633..748R} studied the $BzK$ and DRG fraction 
of submm galaxies with $S_{850\mu\mathrm{m}} \ga 5$\,mJy in 
the GOODS-North field. Out of 11 radio-detected submm galaxies, 
they found that 5 (45$\pm15$\,\%) and 3 ($27\pm13$\,\%) objects satisfy the $BzK$ and 
DRG criteria, respectively. For these submm-bright 
$BzK$ and DRG samples, 2 objects each are too faint to be 
detected at our detection limits. Subtracting these $K$-band 
faint sources would give 
the $BzK$ and DRG fractions of $27\pm 13$\,\% and $9\pm 8$\,\%, 
respectively. On the other hand, \cite{2005ApJ...631L..13D} 
found only one $BzK$-selected galaxy in the list of submm sources 
in \cite{2005MNRAS.358..149P} in the GOODS-North field.
These results are not inconsistent with ours, 
considering the large errors, due to a small sample size and cosmic variance. 

Since the redshift distribution of $BzK$s has a large overlap with that of 
submm galaxies, and both classes of object are 
star-forming galaxies, one could expect 
a large fraction of $BzK$s within the submm galaxy population.
We roughly estimate the expected fraction of $BzK$s in submm galaxy samples as follows, 
based on the results of the redshift survey of radio-detected submm galaxies 
by \cite{2005ApJ...622..772C}. In their sample of radio-detected submm 
galaxies, the fraction of galaxies at $1.4<z<2.5$ is about 45\,\%. This fraction should 
be corrected for the incompleteness of the sample, mainly due to the redshift desert 
at $1.2<z<1.8$. Given the 20\,\% incompleteness at $z\simeq 1.2$--1.8 
estimated by \cite{2005ApJ...622..772C}, 
the fraction of galaxies at $1.4<z<2.5$ may be in the range of 35--45\,\%. 
If $BzK$s are an almost complete sample of star-forming galaxies at $1.4 < z < 2.5$, 
including heavily obscured objects, most of radio-detected submm 
galaxies at $1.4<z<2.5$ may be selected as $BzK$s. In our SHADES/SIRIUS field, 
there are 13 radio-detected SHADES catalogue sources. Therefore, one may expect 
that 4--6 of them are $BzK$s. From the $K$-band magnitude of radio-detected 
submm galaxies of \cite{2004ApJ...616...71S}, we found that $\sim$65\,\% of 
radio-detected submm galaxies are detectable at the detection limit of our 
$K_\mathrm{s}$-band image. 
Thus, we expect that at most $\sim$4 radio-detected submm galaxies 
would be $BzK$s in our sample. This rough estimate is consistent with 
the results above, since we found 3 such radio-detected submm galaxies.
Although the sample is small, this suggests that the $BzK$ selection technique 
is effective even for submm galaxies, i.e. heavily obscured galaxies with extremely 
high SFR. Thus, submm galaxies at $z\sim 2$ may have the largest 
overlap with $BzK$-selected galaxies, when compared to EROs and DRGs. 
In Section 6.2.3, we discuss the implications of this, using 
a radiative transfer model of starburst galaxies. 

The results presented here are hampered by small sample size, partly 
due to the limited overlap between our NIR images and the SHADES 
field. Both of the SHADES fields have been covered by the UKIRT Infrared 
Deep Sky Survey \citep[UKIDSS --][]{2006astro.ph..4426L}. 
In the near future it will therefore be possible to study the 
submm properties of a large sample of NIRGs using the UKIDSS data.

\section{Stacking analysis for submm-faint NIRG{\sevensize s}}
We can try to extract a little more information about the submm galaxies/NIRG 
overlap by looking at the statistical correlations within the images. 
Here we estimate the average 850-$\mu$m flux 
of submm-faint NIRGs and the contribution to the extragalactic background 
light (EBL) from each class of objects. 

We first eliminated the effects of resolved submm sources in our SCUBA maps. 
In the B-map, we excluded regions within $7''$ of the SHADES catalogue 
sources. In the D-map, all the SHADES sources and their associated negative 
off-beams were removed before measuring the flux. For this cleaned D-map, 
we confirmed that random positions give a variance-weighted average flux 
of zero. Although a non-SHADES source for ID1390 is not removed from the 
maps, its effect is negligible. 

In Figures \ref{stacking1}--\ref{stacking3}, 
we show the histogram of measured 
flux at the position of NIRGs in the B-map, compared with that of the map as a whole.
The significance of the difference between the distributions of 
the map and the measured fluxes was estimated with the 
Kolmogorov-Smirnov test. The flux distributions 
of $BzK$s and EROs were found to be different from that of the map
as a whole, with a significance of $2$--$3\,\sigma$ (see Table \ref{tab_stack}). 
For DRGs, we found no significant detection from the 
Kolmogorov-Smirnov test using 56 objects. 
As a comparison, \cite{2005ApJ...632L...9K} 
detected an average flux of $0.74\pm0.24$\,mJy for DRGs (excluding 
discrete sources) with a smaller sample of 30, but with a 
lower noise SCUBA map. A possibly large contamination in our DRG sample, 
owing to the shallow $J$-band data, might cause no significant detection. 

From the B-map, 
we obtained an average flux of $0.52\pm0.19$ and $0.53\pm0.16$\,mJy 
for $BzK$s and EROs, respectively. We measured the average flux of data between 
the first and the third quartiles of the sample, which is a robust 
measure against outliers. The errors are given by the median absolute deviation. 
Table \ref{tab_stack} gives a summary of the stacking analysis using the B-map.  
The noise-weighted average flux of $BzK$s, EROs, and DRGs measured in 
the D-map is $0.64\pm0.16$, $0.50\pm0.13$ and $0.42\pm0.23$ mJy, 
respectively, which are consistent with the results from the B-map. 
The derived average fluxes of $BzK$s and EROs 
are consistent with previous studies in other sky regions 
\citep{2004ApJ...605..645W,2005ApJ...631L..13D}\footnote{Note 
that Daddi et al. (2005) derived 
an average flux of $1.0\pm0.2$\,mJy for `24\,$\mu$m-detected' 
$BzK$s. Including 24\,$\mu$m-undetected objects, this average 
becomes $0.63\pm0.17$\,mJy in the GOODS-North field}. 
In the following discussion, we use the average flux of submm-faint 
$BzK$s and EROs from the B-map.

We next estimate the contribution from each class of object to the EBL 
at 850\,$\mu$m. 
The total flux from individually detected sources is 
added to the estimated total flux from undetected objects. 
The values obtained for the EBL from $BzK$s and EROs are $3.8\pm1.2$ and 
$5.1\pm1.5$\,Jy\,deg$^{-2}$, respectively. 
The measured 850\,$\mu$m EBL is 31\,Jy\,deg$^{-2}$ in 
\cite{1996A&A...308L...5P} and 44\,Jy\,deg$^{-2}$ in 
\cite{1998ApJ...508..123F}, i.e. there is a relatively large uncertainty 
on the absolute level. 
The contribution from 
$BzK$s and EROs are both 10--15\,\% each, depending on the 
adopted value of the EBL. This modest contribution is also suggested by a
comparison of the surface 
density of objects. For example, the surface density of $BzK$s, 
$\sim 1.5$\,arcmin$^{-2}$ is a factor of $\sim 5$ smaller than 
that of $>$0.5\,mJy SCUBA sources, $\sim 7$\,arcmin$^{-2}$ 
\citep{1999ApJ...512L..87B,2002AJ....123.2197C} at which flux density level 
the background is close to complete. 

We also estimate the resolved fraction of the EBL originating from 
a given galaxy population only. The resolved fraction of the EBL from $BzK$s 
is at least $\sim 30$\,\%, which could be 
higher than that found in EROs. The detected sources in our sample 
correspond to $\ga 3.5\,\sigma$ sources when we measure the peak 
flux in the SCUBA map with the noise level of $\sim 2$\,mJy. 
From the 8-mJy survey, which has a similar noise level, 
\cite{2002MNRAS.331..817S} found that $\ga 3.5\,\sigma$ sources 
account for $\simeq 10$\,\% of the EBL. Thus, the submm flux of 
$BzK$s is apparently biased high; i.e. a large fraction of the submm 
fluxes from $BzK$s are found in resolved bright sources. 

Recently, \cite{2006ApJ...647...74W} suggested that the majority of the EBL at 
850\,$\mu$m originates from submm-undetected galaxies at $z\la 1.5$. 
On the other hand, the majority of submm-detected galaxies, i.e. resolved 
sources, lie at $z\ga 1.5$ \citep{2005ApJ...622..772C}. 
This suggests that the resolved fraction of the EBL varies as a function of 
redshift. If we study the EBL only from galaxy populations at $z\ga 1.5$ 
like $BzK$s, the contribution from submm-undetected galaxies to the EBL 
would be significantly reduced. This may explain why the resolved fraction 
of the EBL from $z\sim 2$ $BzK$s is higher than that from EROs typically 
at $z\sim 1$. 


\section{SED analysis of submm-bright NIRG{\sevensize s}}
In this section, we investigate the physical properties of 
submm-bright NIRGs from the observed SEDs. Firstly, we examine
the distribution of submm-bright NIRGs in the $B-z$ vs $z-K$ colour-colour plot, 
i.e. $BzK$ diagram. Specifically, we compare the properties of submm-bright 
NIRGs with those of 24\,$\mu$m-detected NIRGs. 
Secondly, we apply 
an evolutionary SED model of starbursts to submm-bright 
NIRGs for a more detailed study of each object. Also, we  study the 
properties of submm-faint NIRGs from an average SED, which are 
compared with those of submm-bright NIRGs. 

\subsection{Submm-bright NIRGs in the $BzK$ diagram} \label{sec:bzk}
The fraction of $BzK$s in a galaxy 
population would depend on its redshift distribution; i.e. low-$z$ 
galaxy population would have a small fraction of $BzK$s. 
Figure \ref{bzk_obs} shows the 
$BzK$ diagram for all the $K_\mathrm{s}$-detected objects in the 
SXDF/SIRIUS field. 
While NIRGs have a wide range of $B-z$, i.e. $0 \la (B-z) \la 6$, 
submm-bright NIRGs favour the colour range $1 \la (B-z) \la 3$, 
with $BzK \ga -0.2$. 
This is not the case for general 24\,$\mu$m-detected objects, which have 
$0.5 \la (B-z) \la 4$ and a large fraction of non-$BzK$s. 
Unlike 24\,$\mu$m-detected NIRGs, submm-bright NIRGs
disfavour non-$BzK$ EROs and are better matched to 
the $B-z$ colours of $BzK$s. 
This is consistent with the 
redshift distribution of radio-detected submm 
galaxies, having a median of $z\sim 2$ \citep{2005ApJ...622..772C}, 
which overlaps considerably with that of $BzK$s.  
On the other hand, 24\,$\mu$m-detected 
objects include a large number of $z\la 1.5$ galaxies \citep{2005AJ....129.1183R}, 
and have a small $BzK$ fraction. 

Note that 24\,$\mu$m-detected NIRGs include fairly blue $BzK$s 
with $(B-z)\la 1$, unlike submm-bright ones. 
The correlation between $B-z$ and reddening 
for $BzK$-selected galaxies is discussed by \cite{2004ApJ...617..746D},
suggesting that $E(B-V) = 0.25(B-z+0.1)$ for the Calzetti extinction 
law \citep{2000ApJ...533..682C}.
According to this relation, $(B-z)\la 1$ corresponds to $E(B-V)\la 0.3$.
On the other hand, we derive $E(B-V)\simeq 0.5$ for submm-bright 
$BzK$s on average from $\langle B-z \rangle = 2.1$. 
\cite{2005ApJ...633..748R} show that 
UV-selected `BX/BM' galaxies 
occupy a region in the $BzK$ diagram similar to 
$BzK$-selected galaxies with $(B-z) \la 1$. Since these UV-selected 
galaxies are less obscured by dust than submm galaxies 
\citep[e.g.][]{2004ApJ...616...71S}, it is 
expected that most submm-bright NIRGs may avoid the colour 
region of $(B-z)\la 1$.

The other possibilities for blue $B-z$ colours include the presence
of AGN. We performed a cross-correlation between the 24\,$\mu$m-detected 
$BzK$ sample and X-ray sources in the XMM serendipitous source 
catalogue\footnote{The XMM-Newton Serendipitous Source Catalogue, 
version 1.1.0, XMM-Newton Survey Science Centre Consortium, 
XMM-SSC, Leicester, UK (2004)}. As a result, we found that the 
bluest four 24\,$\mu$m-detected $BzK$s with $(B-z) \la 1$ and 
$(z-K) \la 1$ are associated with X-ray sources with an angular 
distance of $\theta < 2''$, while the submm-bright sample have no
such associations. Therefore, bluer colours of 
24\,$\mu$m-detected $BzK$s appear to be better explained by a contribution 
from AGN. 

In summary, the colours of submm-bright NIRGs, 
$1 \la (B-z) \la 3$, indicate that they are 
obscured star-forming galaxies at $z\ga 1.4$, with no obvious 
contribution from AGN to the observed optical-NIR fluxes. 
The nature of submm-bright NIRGs is further discussed below 
using multi-wavelength data and a theoretical SED model.

\subsection{Comparison with SED models}
\subsubsection{SED fitting method}
We analyse the SEDs of submm-bright NIRGs using 
an evolutionary SED model of starbursts of 
\cite{2003MNRAS.340..813T} which has previously been applied to 
submm galaxies in \cite{2004MNRAS.355..424T}. 
In this model, the equations of radiative 
transfer are solved for a spherical geometry with centrally concentrated 
stars and homogeneously distributed dust. 
We use the same model templates as those used in 
\cite{2004MNRAS.355..424T}. We hereafter refer to this model 
as the StarBUrst Radiative Transfer (SBURT) model. 
Here we extend the wavelength range of the SBURT model to the radio by 
assuming the observed correlation between far-IR and radio 
emission \citep{1992ARA&A..30..575C}, 
with $\alpha = -0.75$ and $q = 2.35$, where $\alpha$ is the 
spectral index of radio emission ($S_\nu \propto \nu^\alpha$) 
and $q$ defines the luminosity 
ratio of far-IR to radio emission at 1.49 GHz 
\citep[see][]{1992ARA&A..30..575C}. 

The fitting parameters of the SBURT model 
are the redshift, starburst age and compactness of a 
starburst region ($\Theta$). 
The evolutionary time-scale of the starbursts $t_0$ is assumed 
to be 0.1\,Gyr, which specifies both the gas infall rate and the star formation rate.  
The compactness of starbursts is 
defined by $r = \Theta (M_*/ 10^9 M_\odot)^\frac{1}{2}$ [kpc], where $r$ and $M_*$ are 
the radius and stellar mass of the starburst region, respectively. 
The dust model is chosen 
from the MW, LMC or SMC models taken from \cite{2003PASJ...55..385T}.
Following the results of \cite{2004MNRAS.355..424T}, 
we adopt a top-heavy initial mass function (IMF) 
with a power law index of $x=1.10$ (the Salpeter IMF has $x=1.35$) 
for submm galaxies, which is necessary to reproduce the colour-magnitude 
relation of present-day elliptical galaxies. 
The lower and upper mass limits
of the adopted IMF are 0.1 and 60\,$M_\odot$, respectively. 
This particular choice of the IMF does not affect 
the following results, except for the derived stellar masses. 

The best-fitting SED model is searched for using a $\chi^2$-minimization 
technique from the prepared set of SED models. 
We used all the available fluxes, except for 24\,$\mu$m and radio. 
This is because, in the SBURT model;
1) the contribution from an AGN is not taken into account; 
2) the 24\,$\mu$m flux depends on the details of the 
dust model; and 3) the radio flux is separately calculated by using 
the empirical relation. The adopted 850-$\mu$m fluxes and errors are 
`deboosted values' (i.e. accounting for the effects of flux boosting on a low S/N
threshold sample) taken from Coppin et al. (2006) or 
calculated with the same deboosting algorithm. 
We adopted a minimum flux error of 5\,\%, considering the systematic 
uncertainty of photometry from the different type of instruments covering 
a wide range of wavelength. 
The upper limits on fluxes are taken into account in the 
fitting process, i.e. models exceeding the 5\,$\sigma$ upper 
limits are rejected. 
At rest frame UV wavelengths, the SEDs of heavily obscured starbursts 
like submm galaxies could 
depend on the inhomogeneity of interstellar medium, 
since the effects of photon leakage may dominate the resulting SED 
\cite[e.g.][]{2003MNRAS.340..813T}. Also, the absorption by the 
intergalactic medium is important in the rest frame below $1216$\,\AA, which depends 
on a particular line of sight to each galaxy. Considering these 
uncertainties, we quadratically added an additional 20\% error for 
data at rest-frame UV wavelengths below 4000\,\AA.

\subsubsection{Results of SED fitting}
In Figure \ref{sed}, we show the best-fitting models for 
submm-bright NIRGs. The original and radio-extended SBURT 
models are depicted with solid and dotted lines, respectively.
The fitting and derived model parameters are summarised in Table 
\ref{tab_takagi}. 

For ID912, we found that the value of minimum $\chi^2$ reduced drastically 
when we excluded $H$-band data from the fitting analysis. This may indicate 
that the $H$-band photometry could have a large systematic error. The upper 
limit in the $J$-band suggests that this galaxy could be very red in the observed 
NIR colours. The noise level in the $H$-band may be too high to detect this galaxy. 
We checked the images of ID912 and found that photometry of this galaxy could be 
affected by a nearby object. Hence, we adopt the best fitting model 
for ID912 without $H$-band data. 

We find a significant underestimate of 850\,$\mu$m flux for ID445 which has
a very red $z'-K_\mathrm{s}$ colour, although the resulting $\chi^2$ value indicates that 
the best-fitting model is statistically acceptable. Since the predicted radio flux is also well 
below the observed flux, the best-fitting 
model might be rejected with more accurate measurements in the submm. 
For this source, we may particularly need a more complicated multi-component model of 
starbursts, in which young heavily obscured molecular clouds are treated 
separately \citep[e.g.][]{1998ApJ...509..103S}. 

The SED fitting suggests that submm-bright NIRGs could lie at the 
typical redshift range of submm galaxies \citep{2005ApJ...622..772C}. 
In order to show the fitting error on photometric redshifts $z_\mathrm{phot}$,
we show contour plots of $\Delta \chi^2$ projected onto 
the age-redshift plane in Figure \ref{error} for models with the 
best dust model for each galaxy. We note that 
the derived $z_\mathrm{phot}$ and the far-IR--radio relation 
reproduce the observed radio flux well, except for ID445. 

At 24\,$\mu$m, we find that the observed fluxes are typically higher 
than the model predictions, while the model flux of ID912 is consistent 
with the observed 24\,$\mu$m flux. 
We regard the model flux at 24\,$\mu$m as the lower limit for 
the following reasons: 1) lack of AGN contribution in the model;
and 2) the non-MW dust models are preferentially selected from the 
featureless SED at rest-frame UV, which 
may have a lower MIR emissivity than actual dust grains in submm
galaxies. Specifically, the lack of AGN could be important for 
submm galaxies, in which the AGN activity to some extent 
is already known \citep{2005ApJ...632..736A}. 
Recently, \cite{2007arXiv0705.2832D} found a galaxy population 
at $z\sim 2$ which shows a distinct excess of flux at 24\,$\mu$m, 
compared to that expected from the SFR estimated at other wavelengths. 
This MIR excess is attributed to Compton-thick AGNs, as a result of 
stacking analysis of deep X-ray images. 
Submm-bright NIRGs showing a clear excess of flux at 24\,$\mu$m 
(ID300 and 1390), might be such MIR-excess galaxies. 

Submm-bright NIRGs are found to be massive, 
with the stellar masses of $5\times 10^{10}$--$10^{11}$~M$_\odot$. 
The derived bolometric luminosity 
of $3\times 10^{12}$--$10^{13} $L$_\odot$ (excluding ID445) 
corresponds to the star formation rate of $\sim 300$--1000~M$_\odot\,$yr$^{-1}$ for 
the adopted top-heavy IMF. For the Salpeter IMF, the stellar mass and 
the SFR could be even a few times higher than we derived. 

Following \cite{2004MNRAS.355..424T}, we predict the present-day
colours and absolute magnitudes of submm-bright NIRGs. Since we assume 
that the effects of star formation after the observed epoch is 
negligible, the derived present-day 
colours and luminosities are both lower limits. 
These lower limits could be close to reasonable values for ID300 
and 1390, which seem to be well evolved starbursts with 
$t/t_0 \ga 2$ and $M_\mathrm{star} > M_\mathrm{gas}$. 
For these galaxies, we predict absolute $V$-band magnitudes of 
$M_V = -20.7$ to $-20.1$, while submm-faint $BzK$s would have 
$M_V =-19.4$ on average. Since we use a top-heavy IMF, this magnitude 
is $\sim 1$ mag fainter than those by the Salpeter IMF. 
We find that the predicted rest-frame $U-V$ and $M_V$ are consistent with 
the colour-magnitude relation of elliptical galaxies. This is 
not the case for the Salpeter IMF \citep[see also][]{2004MNRAS.355..424T}. 

\subsubsection{Colour evolution of the SBURT model}
In Section 4, we mentioned the possibility that submm galaxies have a 
much larger overlap with $BzK$-selected galaxies, compared with EROs and DRGs. 
In Figure \ref{bzk_col}, we show the $BzK$, $R-K$, and $J-K$ 
colours as a function of starburst age at $z=2$, i.e. a typical 
redshift for submm galaxies. 
We choose $\Theta = 0.7$--1.4, 
which results in a good match with the observed SED variation of submm 
galaxies for the SMC dust model \cite[see][]{2005MNRAS.357..165T}. 
The SBURT models satisfy $BzK\ga -0.2$ for a wide range of model 
parameters, i.e. starburst age and compactness of the starburst 
region (or optical depth). On the other hand, $R-K>3.35$ is 
only satisfied with old models having $t/t_0 \ga 3$ for a wide range of $\Theta$. 
This is also true for DRGs, except for a small fraction of models. 
Therefore, in the SBURT model, SEDs of stellar populations 
need to be intrinsically red to reproduce the colours of EROs and DRGs. 
Thus, this model predicts the largest overlap between $BzK$s 
and submm galaxies, among NIRGs. 
If we assume the limit on 
starburst age of $t/t_0 \sim 6$ \citep{2004MNRAS.355..424T}, 
we expect the number of $BzK$s in submm galaxies to be about 
twice that of EROs, given that the selection of $BzK$s and EROs are 
not sensitive to the compactness $\Theta$. 
This prediction could be confirmed by using larger samples of submm-bright 
NIRGs.

\subsubsection{Why are some $BzK$s bright in submm?}
In this study, we found that only a handful of NIRGs are bright in the submm. 
So, are submm-bright ones experiencing a special luminous evolutionary 
phase or simply more massive than submm-faint ones? 
If the former is the case, the 
duty cycle of submm-bright $BzK$s may be estimated from the fraction of 
submm-bright NIRGs. In the following discussion, we focus only on $BzK$s, since EROs 
and DRGs could be contaminated by passively evolving galaxies. 
We found that only 3\,\% of $BzK$s are bright in submm. This indicates that 
the duty cycle of submm-bright $BzK$s is only $<$10\,\% of the star-forming phase, 
even if we consider the incompleteness of the SHADES survey 
(Coppin et al. 2006). Thus, we might be observing a very sharp 
peak in the star formation activity of $BzK$s \citep[e.g. see also][]{2006ApJ...637L...5D}. 
If this is the case, the SEDs of submm-bright $BzK$s may be systematically 
different from those of submm-faint ones.

In order to address this question, we derived the average SED of submm-faint 
$BzK$s with $1.5< (B-z) < 2.5$, i.e. having similar colour to 
submm-bright ones. 
In Figure \ref{sed_ave}, we show the average SED of 65 
submm-faint $BzK$s with $1.5< (B-z) < 2.5$. At 850\,$\mu$m, 
we adopted the average flux of $BzK$s derived from our 
stacking analysis. The average flux of submm-faint $BzK$s 
in the radio were also derived with a stacking analysis, and 
found to be $f_\mathrm{1.4GHz} = 5.7\pm 1.0$\,$\mu$Jy.  
A representative model\footnote{We call this model not the `best' 
but `representative', since the variance of the average SED is too large to 
specify one particular model as the best one.} for the average SED was 
sought with the same method used for individual submm-bright galaxies. 
For the flux error at optical-NIR bands, we adopted the standard 
deviation of the sample at each photometric band. Since this error is 
rather large, we fixed the redshift to the average redshift of $BzK$s, 
$\langle z \rangle=$1.9 \citep{2004ApJ...600L.127D}, in order to 
constrain the parameter space. We show the model thus obtained 
in Figure \ref{sed_ave}. 
The model parameters are tabulated in Table \ref{tab_takagi}. 

Although the uncertainty is large, the average SED thus derived is not 
very much different from those of submm-bright $BzK$s, and reasonably 
explained by the the SBURT model, i.e. a starburst model, not mild 
evolutionary models for quiescent galaxies. Therefore, from the SED
analysis, we found no clear signatures that the evolutionary phase 
of submm-bright $BzK$s is substantially different from that of submm-faint ones. 
Compared to submm-bright galaxies, a difference in the model 
parameters may be found in the mass scale, i.e. 
submm-faint ones are less massive. Since the representative model 
is an old model of a starburst, the estimated mass is close to the 
upper limit, owing to a higher mass-to-light ratio compared with younger
models. From an oldest model which reproduces the average SED, 
we derive the upper limit on the stellar mass as 
$M_\mathrm{star}  \la 5\times 10^{10}$~M$_\odot$. 
Compared to the oldest submm-bright 
NIRG in our sample, ID1390 (excluding ID445), the stellar mass of 
submm-faint $BzK$s could be less than half of ID1390.
This may suggest that submm-bright and faint $BzK$s evolve into 
galaxies with different stellar masses, rather than being
galaxies of a similar stellar mass but in different evolutionary stages. 
We however caution that the sample size of submm-bright $BzK$s are 
too small to derive firm conclusions on their average physical properties, 
such as stellar mass. We need a larger sample of submm-bright $BzK$s, and 
spectroscopic redshifts for more secure analyses. 
Also note that the mass estimate of the SBURT model may suffer
from systematic effects, owing to the assumed simple star/dust geometry.


\section{Summary}
We have investigated the submm properties of the following 
classes of 
NIR-selected galaxies: $BzK$-selected star-forming galaxies; DRGs; and EROs. 
We utilised a 93$\,$arcmin$^2$ sub-region of the SHADES SXDF 850$\mu m$
map which with has already been imaged in the NIR with the 
SIRIUS camera on the UH\,2.2m telescope.

Using two SCUBA 850-$\mu$m maps (the SHADES B-map and D-map) 
produced by two different groups 
within the SHADES consortium, we detected 6 NIRGs above 3$\sigma$. 
Four submm-detected NIRGs out of six are also detected both at 24\,$\mu$m 
and in the radio. This suggests that these 4 submm-detected NIRGs 
are genuine submm-bright galaxies. 
These submm-bright NIRGs are all $BzK$-selected galaxies, 
except for ID1390 whose $BzK$ colour is however close to the 
selection boundary.  In other words, no EROs and DRGs are found to be
submm-bright if they are clearly non-$BzK$s. We made a rough estimate 
of the number of $BzK$s in radio-detected submm galaxies, assuming 
that submm galaxies at $1.4 < z <2.5$ satisfy the selection criteria of 
$BzK$s. Although the sample is small, this estimate is consistent with 
our result. This may indicate that most submm galaxies at $1.4 <z<2.5$ 
could be $BzK$s. 

Two submm-detected NIRGs satisfy all the selection criteria we 
adopted, i.e. they are extremely red $BzK$-selected galaxies. 
Although these extremely red $BzK$s are rare, the fraction of 
submm galaxies in them could be high (up to $20$\,\%), compared 
with the other colour-selected optical-NIR galaxy populations. 

We performed stacking analyses with our SCUBA 850-$\mu$m maps, 
in order to derive the average flux of submm-faint NIRGs. We derived
$0.52\pm 0.19$ ($0.64\pm 0.16$) mJy and $0.53\pm0.16$ ($0.50\pm 0.13$) mJy
from the B-map (D-map) for $BzK$s and EROs, respectively, while 
we found no significant signal from DRGs in either map. 
The contribution from $BzK$s and EROs to the EBL at 850\,$\mu$m 
is about 10--15\,\%. Focusing on the EBL only from $BzK$s, 
we found that $\ga 30$\,\% of the EBL 
from $BzK$s is resolved in our SCUBA map. This is higher than 
that for EROs and submm sources as a whole. This might be expected 
if the majority of the EBL originates from submm-undetected galaxies at $z<1.5$, as 
suggested by \cite{2006ApJ...647...74W}. For galaxies at $z\sim 2$, 
the fraction of submm flux in resolved sources could be 
higher than that in low-$z$ galaxies. 

We have also fitted SED models of starbursts 
to each of the submm-bright NIRG (mostly $BzK$s) 
and to an average SED of submm-faint $BzK$s  
derived from galaxies with $1.5 < (B-z) < 2.5$, i.e. similar colours to the submm-bright ones. 
Submm-bright NIRGs are found to lie at the typical redshifts of submm galaxies 
and have stellar masses of $5\times 10^{10}$--$10^{11}$~M$_\odot$ 
with a Salpeter-like, but slightly top-heavy IMF
and bolometric luminosity 
of $3\times 10^{12}$--$10^{13}~$L$_\odot$. 
From the average SED of submm-faint $BzK$s, we found no clear signature that 
the evolutionary phase of submm-bright $BzK$s are substantially different from that of 
submm-faint ones, as suggested by the small number of submm-bright $BzK$s. 
On the other hand, submm-faint 
$BzK$s are likely to be less massive, with the stellar mass below 
$\sim 5\times 10^{10}~$M$_\odot$. 

The results presented here are clearly limited by small sample size. 
Nevertheless, our study can still be considered 
useful for investigating
the physical relationships between NIR-selected and submm-selected massive 
galaxies. A large sample of submm-bright NIRGs from currently ongoing and 
future muti-wavelength surveys, including SHADES and UKIDSS will play 
an important role on the study of massive galaxy formation and evolution.

\section*{Acknowledgments}
We thank all the members of the SHADES consortium, and acknowledge 
continuous support from the staff of the JCMT. We also thank all the 
members of the SIRIUS team and acknowledge the SWIRE project team. 
We thank E.~Daddi for useful discussions
on the $BzK$-selection technique. TT acknowledges the Japan Society for 
the Promotion of Science (JSPS -- PD fellow, No.\ 18$\cdot$7747). 
TT and SS acknowledge support by the Particle Physics and 
Astronomy Research Council under grant number PPA/G/S/2001/00120/2.
IRS acknowledges support from the Royal Society. 
This work is partly supported by a Grant-in-Aid for Scientific 
Research (No. 14540220) by the Japanese Ministry of Education, Culture, 
Sports, Science and Technology. 
KC and AP acknowledge the National Science and Engineering 
Research Council of Canada (NSERC). 
IA and DHH acknowledge partial support from Conacyt grants 
39548-F and 39953-F. 
The JCMT is operated by the 
Joint Astronomy Centre on behalf of the UK Particle Physics and 
Astronomy Research Council, the Canadian National Research Council
and the Netherlands Organization for Scientific Research.

\begin{table*}
\begin{minipage}{100mm}
\caption{Statistics of NIR-selected galaxies (NIRGs)}
\label{tab_stat}
\begin{tabular}{@{}lccc}
\hline
Population  & Area$^a$ &  Number & Surface density \\
  & $K_\mathrm{s}$ or $J$ &  &  [arcmin$^{-2}$] \\
\hline 
$K_\mathrm{s}$-detected object & $K_\mathrm{s}$    & 1308 (992)$^b$  &  11.5  \\
Star                  & $K_\mathrm{s}$   &  95  (64) &  0.83   \\
NIRGs                 & $J$ &   307 (245) & 3.14 \\
ERO                   & $K_\mathrm{s}$ &   249  (201) & 2.18 \\
DRG                   & $J$ &  84  (67)   & 1.09 \\
ERO and non-DRG       & $J$ &   105 (84) &  1.36   \\
DRG and non-ERO       & $J$ &   38 (28) &  0.49   \\
ERO and DRG           & $J$ &   46 (39) &  0.60   \\
$BzK$                 & $K_\mathrm{s}$   &  168 (132) &  1.47   \\
$BzK$ and ERO and non-DRG & $J$& 17 (14) &  0.22  \\
$BzK$ and DRG and non-ERO & $J$& 14 (8)&  0.18  \\
$BzK$ and ERO and DRG   & $J$& 19 (16)&  0.25  \\
\hline 
\end{tabular}
\medskip
a) Image ($K_\mathrm{s}$- or $J$-band area) for which the 
number of objects are derived.
b) Total number of objects, with the number within the SHADES 
area given in parentheses.
\end{minipage}
\end{table*}

\begin{table*}
\begin{minipage}{120mm}
\caption{Submm properties of submm-bright NIR-selected galaxies}
\label{tab_obs1}
\begin{tabular}{@{}lccccccccccc}
\hline
Source  & 
Map$^a$ & 
$S/N^b$ &
$\theta_{850 \mu \mathrm{m}}^c$ & 
$S_{850 \mu \mathrm{m}}^d$ & 
$S_{24\mu\mathrm{m}}$
&
$S_{1.4 \mathrm{GHz}} $ &
SHADES ID$^e$
\\
   &
   &
   &
[arcsec] & 
[mJy] &  
[$\mu$Jy]  &
[$\mu$Jy]  &
[SXDF]   \\
\hline  
  300  &B,D&3.1  &4.5  &   5.7$\pm$   2.1 & 546$\pm$27 & 27.8$\pm$ 7.0  & 850.30 \\
  445  &B  &3.3  &2.9  &   5.6$\pm$   2.1 & 357$\pm$24 & 250.$\pm$ 7.4 & 850.27 \\
  912  &B,D&4.1  &2.1  &   4.4$\pm$   1.8 & 511$\pm$27 & 145.$\pm$ 7.7 &850.4 \\
 1390  &B,D&3.4  &1.8  &   4.0$\pm$   2.8 & 446$\pm$29 & 29.6$\pm$ 7.5 & (850.62)$^f$ \\
\hline  
\multicolumn{8}{c}{Unconfirmed detections} \\
\hline  
  718  &B,D&3.3  &2.3  &   4.0$\pm$   2.1 & $<450$     & $<35$ &  850.70 \\
 1133  &D  &3.0  &6.6  &   3.0$\pm$   2.1 & $<450$     & $<40$ & 850.77\\ 
\hline  
\end{tabular}
\medskip
Notes: The upper limits correspond to 5\,$\sigma$. 
a) SHADES map in which NIRGs are detected
b) $S/N$ at the position of NIRGs in the SHADES map. We adopt 
values from the SHADES B-map if NIRGs are detected in both B and D maps. 
c) The angular separation between submm and NIR position. Submm 
positions are taken from the SHADES source catalogue. For ID1390, 
we adopt the submm position from the SHADES B-map. 
d) Deboosted 850-$\mu$m flux densities from the SHADES source catalogue, 
using the algorithm of \cite{2005MNRAS.357.1022C}. 
e) Source names in the SHADES source catalogue. 
f) Source name from a preliminary SHADES source catalogue.
\end{minipage}
\end{table*}


\begin{table*}
\begin{minipage}{165mm}
\caption{Optical -- NIR properties of submm-bright NIR-selected galaxies}
\label{tab_obs2}
\begin{tabular}{@{}lcccccccccccc}
\hline
Source  & R.A. & Dec. & $K_\mathrm{s}$ &  $R-K_\mathrm{s}$ & $J-K_\mathrm{s}$ & $B-z'$  & $z'-K_\mathrm{s}$ & \multicolumn{4}{c}{Notes}\\
  &[J2000] & [J2000] & \multicolumn{5}{c}{[AB mag]} & $BzK$ & ERO &  DRG & 24\,$\mu$m & Radio \\
\hline  
 300  & 2 17 40.01 & -05 01 15.7  & 21.14  &  3.43  &  1.56  &  2.05  &  2.61 & Y & Y & Y & Y& Y  \\
 445  & 2 18 07.92 & -05 01 45.7  & 22.00  &  4.39  &  $>1.40$  & 1.75  &  3.21  &Y&Y&Y&Y&Y \\
 912  & 2 17 38.67 & -05 03 39.4  & 21.64  &  2.59  &  ...   &  1.20  &  2.05 &Y& N & ... &Y&Y  \\
1390  & 2 18 07.74  & -05 06 10.5  & 21.13  &  3.67  &  2.02  & 2.52  &  2.56 & (N)$^a$&Y&Y&Y&Y \\
\hline  
\end{tabular}
\medskip
Notes: A `Y' ('N') in the last five columns indicate that the galaxy is (not)
$BzK$-selected galaxy, ERO, DRG, 24\,$\mu$m-detected and radio-detected. 
a) After the colour correction on $B-z$ and $z-K$, 
this source becomes a non-$BzK$-selected galaxy, as shown in Figure \ref{bzk_obs}. 
This correction is adopted to match the observed colour sequence of stars 
to that of \cite{2004ApJ...617..746D}. 
\end{minipage}
\end{table*}

\begin{table*}
\begin{minipage}{110mm}
\caption{Summary of stacking analysis at 850\,$\mu$m}
\label{tab_stack}
\begin{tabular}{@{}lccccccccccc}
\hline
Class     &  Number$^a$  &   $\langle S_{850\mu \mathrm{m}} \rangle^b$ & K-S Prob.$^c$ &
        \% Resolved flux$^d$ &  E.B.L.$^e$ \\
        &   & [mJy] & [\%] & [\%] & [Jy\,deg$^{-2}$]  \\
\hline  
$BzK$s  & 112  &$0.52\pm0.19$ & 5.91 &  $32\pm18$   & $3.8\pm1.2$\\
EROs    & 178  &$0.53\pm0.16$ & 1.70 &  $18\pm9$ & $5.1\pm1.5$   \\
DRGs    & 56   &$0.30\pm0.28$ & 45.91 &  ... & ... \\
\hline
\end{tabular}
\medskip
a) The number of objects used for the stacking analysis.
b) The average 850-$\mu$m flux of non-detected objects, which are $>7''$ 
away from individual sources.
c) The probability from the Kolmogorov-Smirnov test.
that the flux distribution of objects are derived from the same sample. 
d) The fraction of the flux density in individually detected sources. 
e) The extragalactic background light at 850\,$\mu$m from each class of objects.
\end{minipage}
\end{table*}

\begin{table*}
\begin{minipage}{150mm}
\caption{Summary of the SED fitting with the SBURT model}
\label{tab_takagi}
\begin{tabular}{@{}lccccccccccccc}
\hline
Source  & $\chi^2/\nu^a$ & $z_\mathrm{phot}$ & Age$^b$ & $\Theta^c$ & Ext.$^d$  & $\log M_\mathrm{star}$ & $\log M_\mathrm{gas}$ 
      & $\log L_\mathrm{bol}^e$ & $A_V$ & $\log \mathrm{SFR}$ & $M_V^f$ & $(U-V)^f$ \\
        &    &              &  [Gyr]        &         &         & [M$_\odot$] &  [M$_\odot$]   
            & [L$_\odot$]  &    [mag]       &   [M$_\odot$ yr$^{-1}$] & [mag] & [mag] \\%
\hline  
 300 & 0.91  &2.8 &  0.2 &  2.0 &  SMC &10.8 &10.7 &12.7 & 1.3 & 2.7 &  $-$20.1 & 1.3 \\ 
445 &  1.47 &  2.1 &   0.5 &   1.0 &  LMC &  10.9 & 10.0 &  12.1 &  2.3 &  2.0 &   $-$20.3 &  1.5  \\
 912 & 0.47 &2.4 &  0.07 &  1.4 &  LMC &10.7 &11.1 &13.1 & 3.1 & 3.2 &  $-$20.1 & 0.9  \\
1390 &2.45  &2.4 &  0.4 &  1.4 &  LMC &11.1 &10.4 &12.5 & 1.5 & 2.4 &  $-$20.7 & 1.4 \\ 
\hline
$\langle BzK$s$\rangle^g$ & ... &
 1.9$^h$ &     0.4 &     1.6 &  LMC &   10.6 &    9.8 &   11.9 &    1.1 &    1.8 &  $-$19.4 &    1.4   \\
\hline
\end{tabular}
\medskip
a) $\chi^2$ divided by the degree of freedom $\nu$.
b) Starburst age with the evolutionary time scale of $t_0$ = 0.1\,Gyr.
c) Compactness parameter of starburst region.
d) Extinction curve used.
e) Bolometric luminosity.
f) Predicted present-day $V$-band magnitude and $U-V$ (Vega), 
assuming the passive evolution after the observed epoch. 
g) Model parameters for an average SED of submm-faint $BzK$s with $1.5< (B-z) < 2.5$
h) Assumed redshift for the SED fitting.
\end{minipage}
\end{table*}



\begin{figure*}
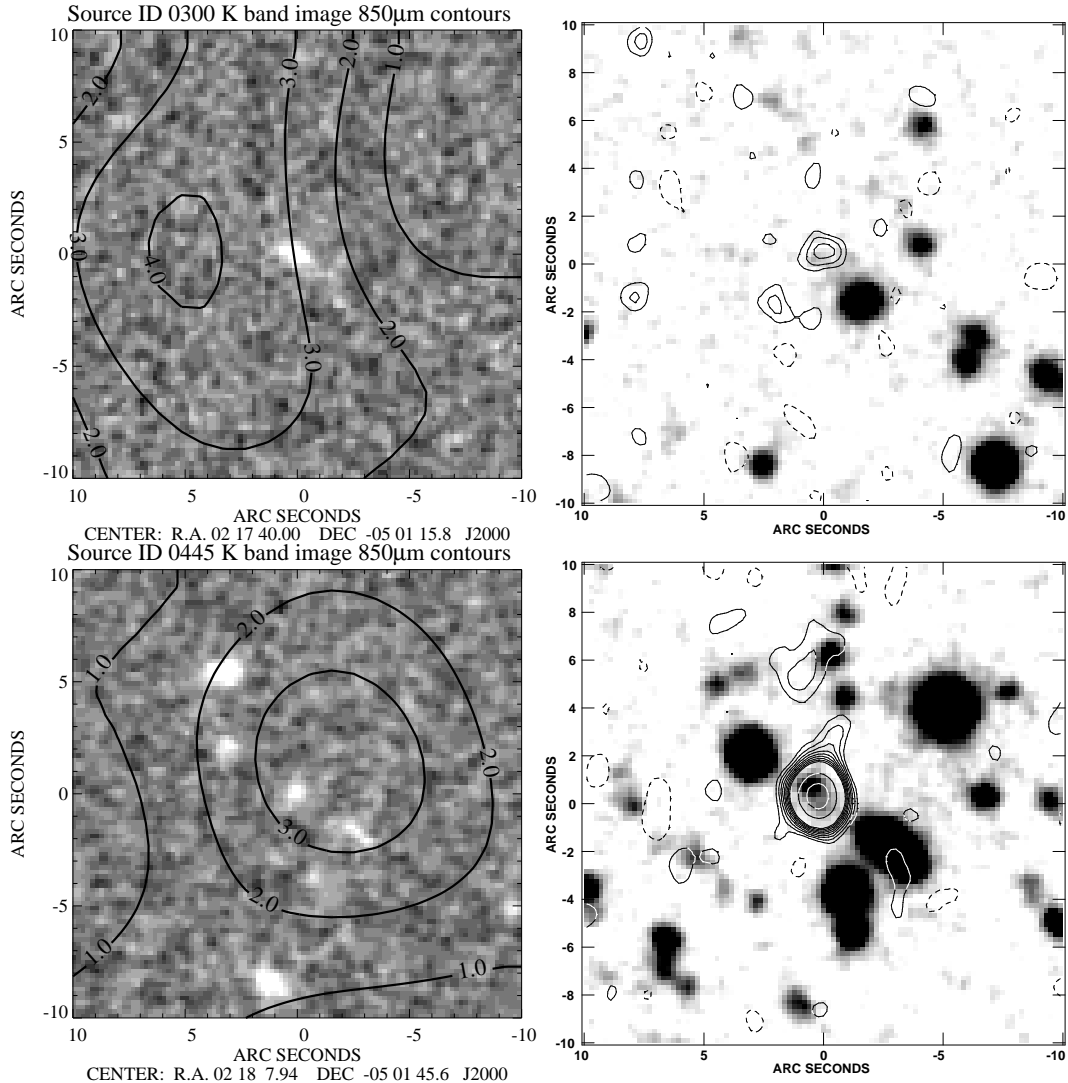

  \resizebox{7cm}{!}{\includegraphics{figure/plot_idsK_0300.epsi}}
  \resizebox{7cm}{!}{\includegraphics{figure/ID300.PS}}
  \resizebox{7cm}{!}{\includegraphics{figure/plot_idsK_0445.epsi}}
  \resizebox{7cm}{!}{\includegraphics{figure/ID445.PS}}
 \caption{[Left column] Contours of SCUBA 850-$\mu$m for submm-bright NIR-selected 
galaxies. The corresponding $S/N$ ratios at 850\,$\mu$m 
are indicated at each contour line. 
The underlying images are $K_\mathrm{s}$-band with 20$^{\prime\prime}$ on a side. 
The names of the sources are indicated at the top of each image. 
[Right column] Contour maps of radio (1.4 GHz) on negative images 
at $R$-band for the same regions as shown in the left panels. 
Contour lines are shown at $-2$, 2, 3, 4, ... ,10, 20, ... 100\,$\times \sigma$. 
Positive (negative) contours are indicated with solid (dashed) lines.
}
 \label{submm}
\end{figure*}

\addtocounter{figure}{-1}
\begin{figure*}
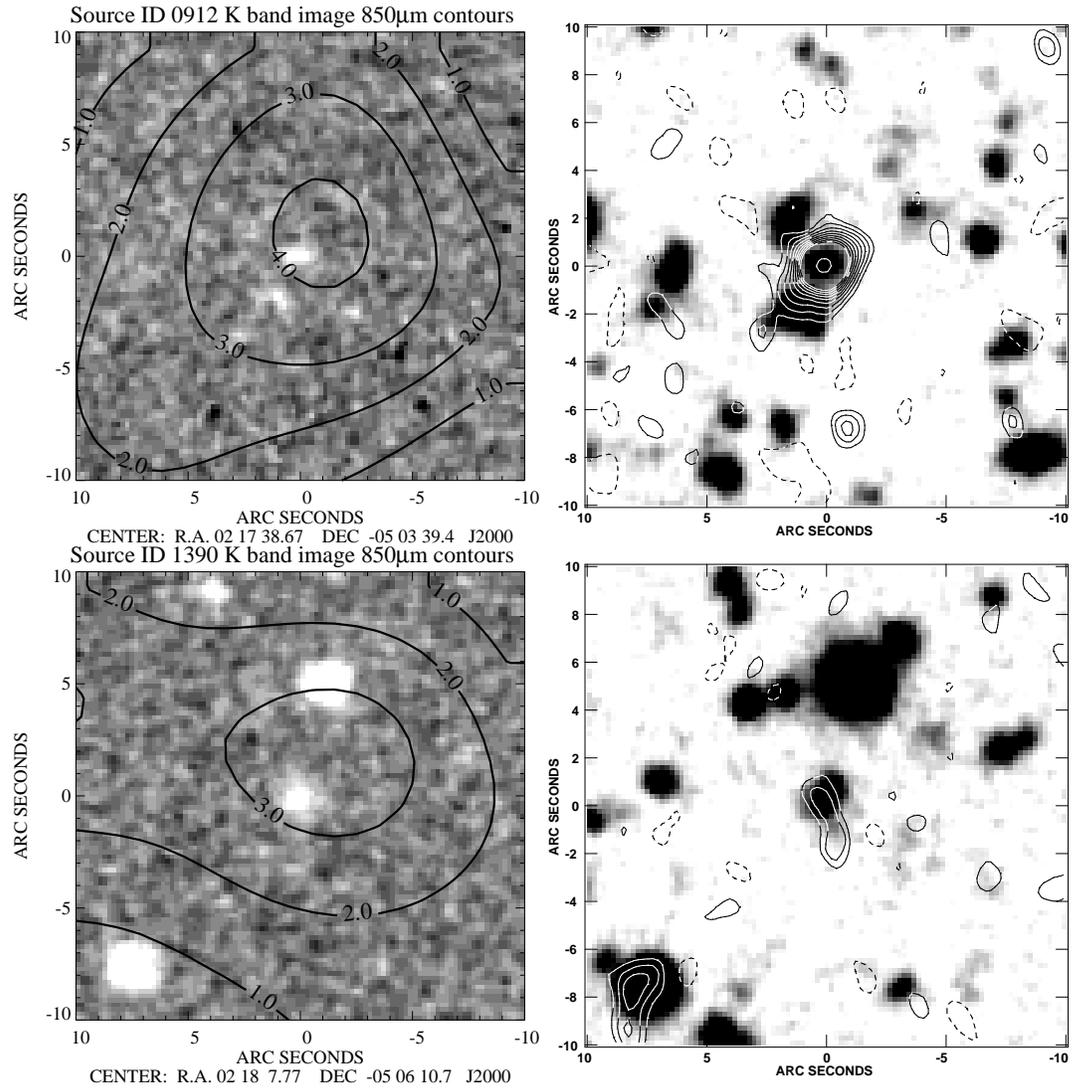

  \resizebox{7cm}{!}{\includegraphics{figure/plot_idsK_0912.epsi}}
  \resizebox{7cm}{!}{\includegraphics{figure/ID912.PS}}
  \resizebox{7cm}{!}{\includegraphics{figure/plot_idsK_1390.epsi}}
  \resizebox{7cm}{!}{\includegraphics{figure/ID1390.PS}}
 \caption{
 {\it - continued.}
}
 \label{submm}
\end{figure*}

  \begin{figure*}
  \resizebox{7cm}{!}{\includegraphics{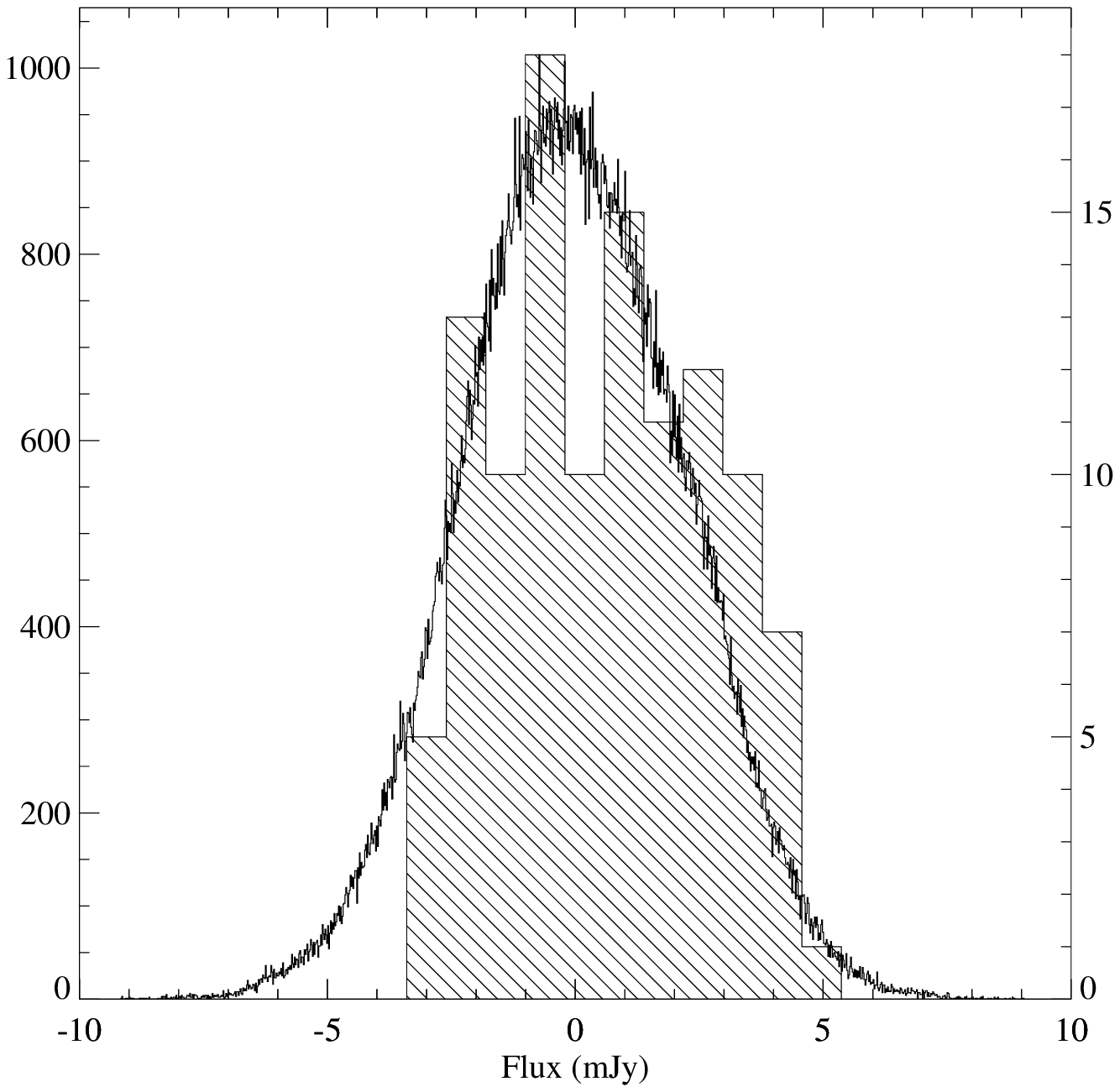}}
 \caption{Histograms 
of the 850-$\mu$m flux at the position of 
$BzK$s (shaded; right axis), 
compared to the histograms for the unmasked 
regions of the SCUBA map as a whole (open; left axis). 
}
 \label{stacking1}
\end{figure*}

  \begin{figure*}
  \resizebox{7cm}{!}{\includegraphics{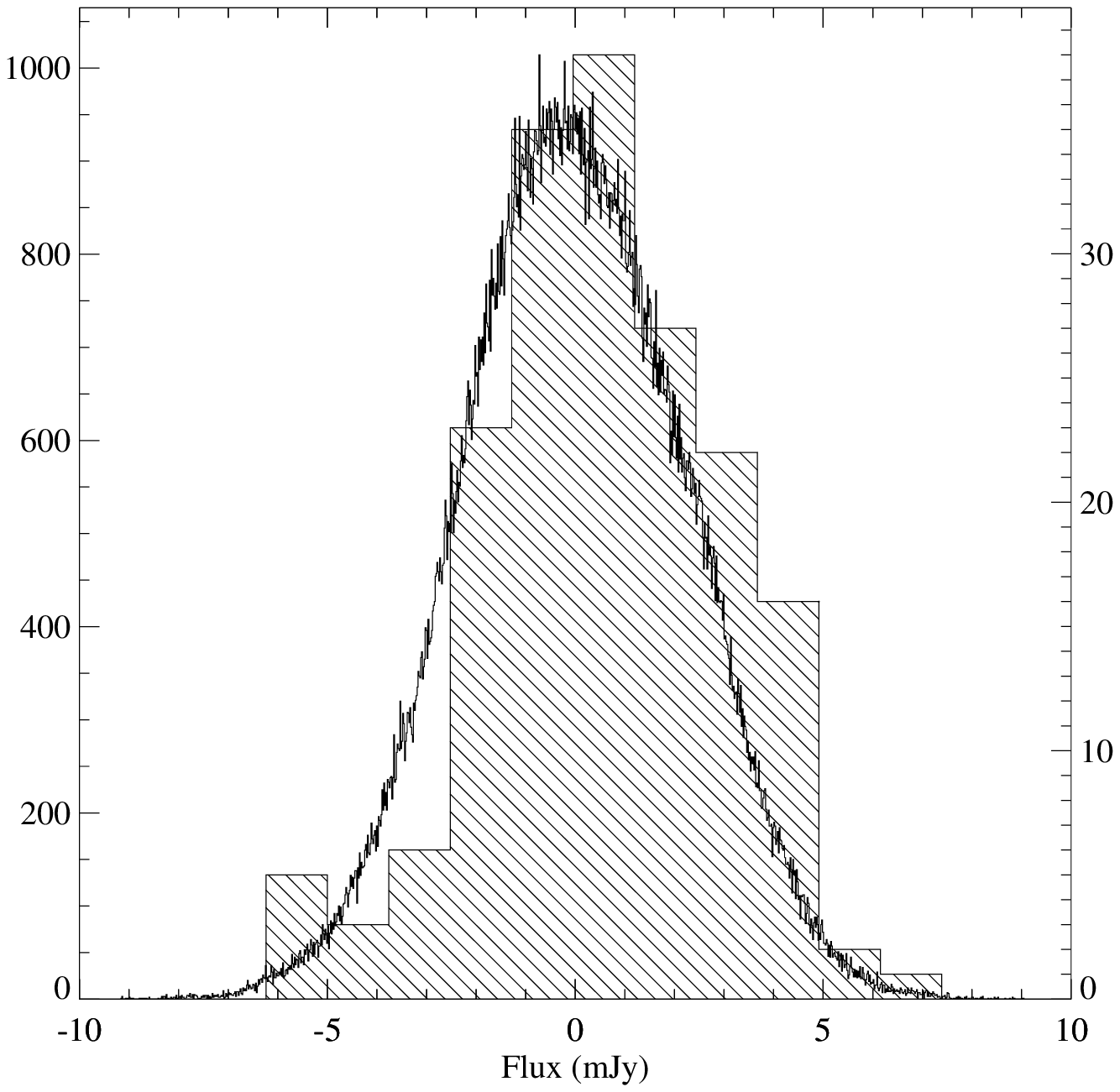}}
 \caption{Same as Figure \ref{stacking1}, but for EROs.
}
 \label{stacking2}
\end{figure*}

  \begin{figure*}
  \resizebox{7cm}{!}{\includegraphics{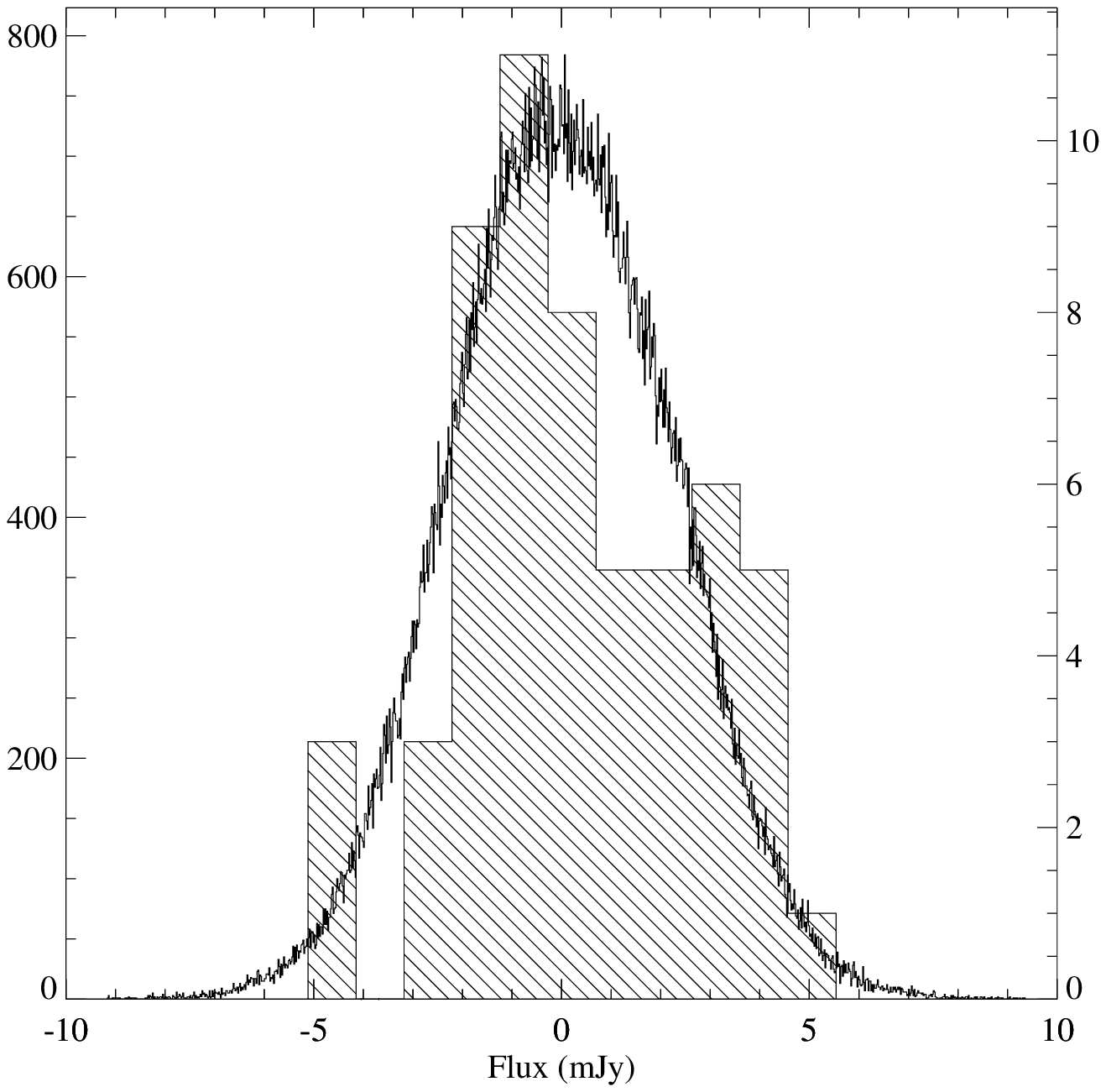}}
 \caption{Same as Figure \ref{stacking1}, but for DRGs.
}
 \label{stacking3}
\end{figure*}


  \begin{figure*}
  \resizebox{\hsize}{!}{\includegraphics{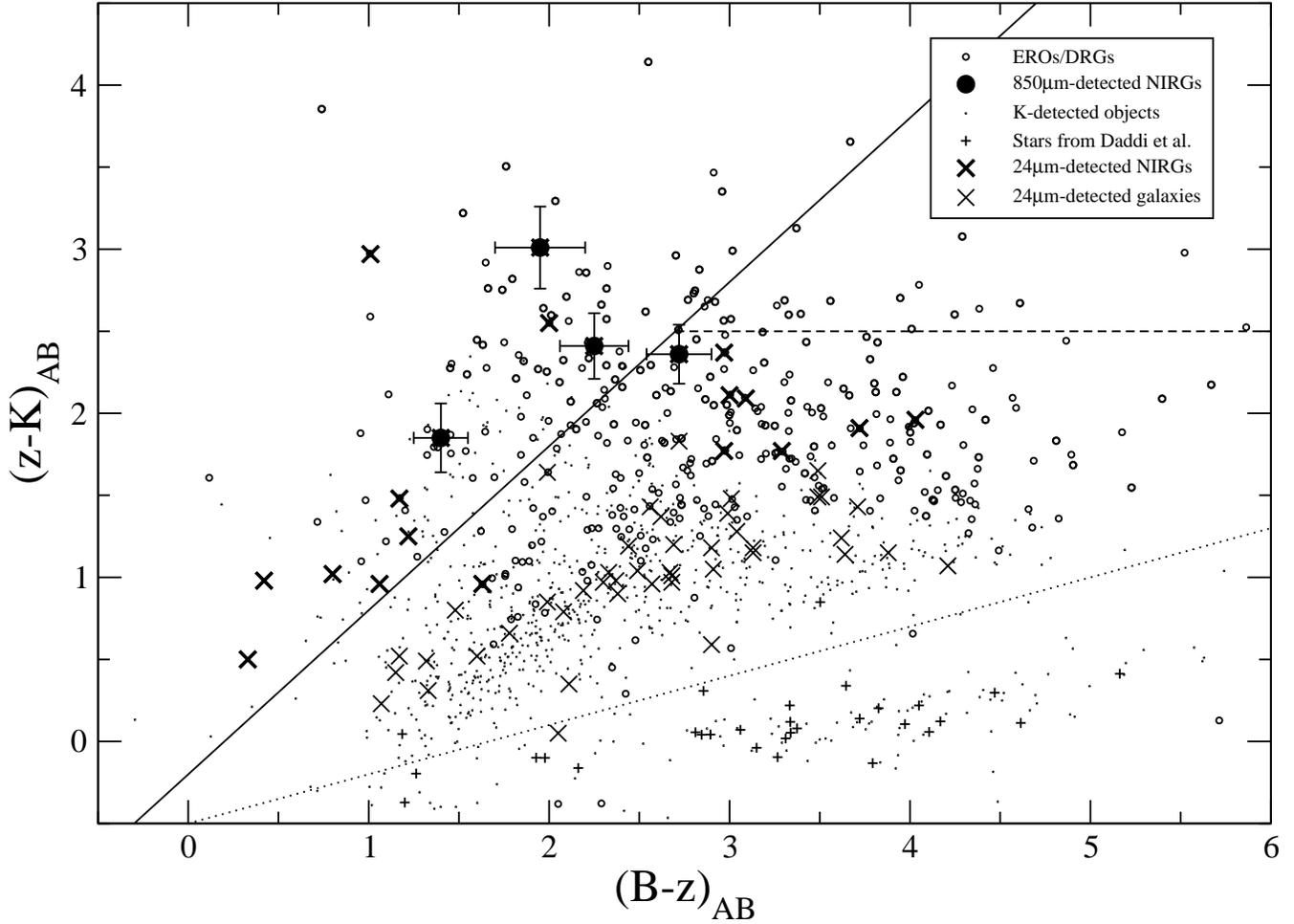}}
 \caption{
The $BzK$ diagram for $K_\mathrm{s}$-detected sources in the SXDF/SIRIUS field. 
Solid circles indicate submm-bright NIRGs. 
The photometric errors include those of colour corrections 
on $B-z$ and $z-K$. 
Both EROs and DRGs are indicated with small circles. 
Large thick/thin crosses are for 
NIR-selected/$K_\mathrm{s}$-detected galaxies which are detected 
at 24\,$\mu$m with SWIRE. 
Small pluses indicate stars from Daddi's catalogue. 
We depict the $BzK$-selection criterion with a solid line. 
Dashed and dotted lines are boundaries for selecting passively evolving 
galaxies and stars, respectively. 
}
 \label{bzk_obs}
\end{figure*}

\begin{figure*}
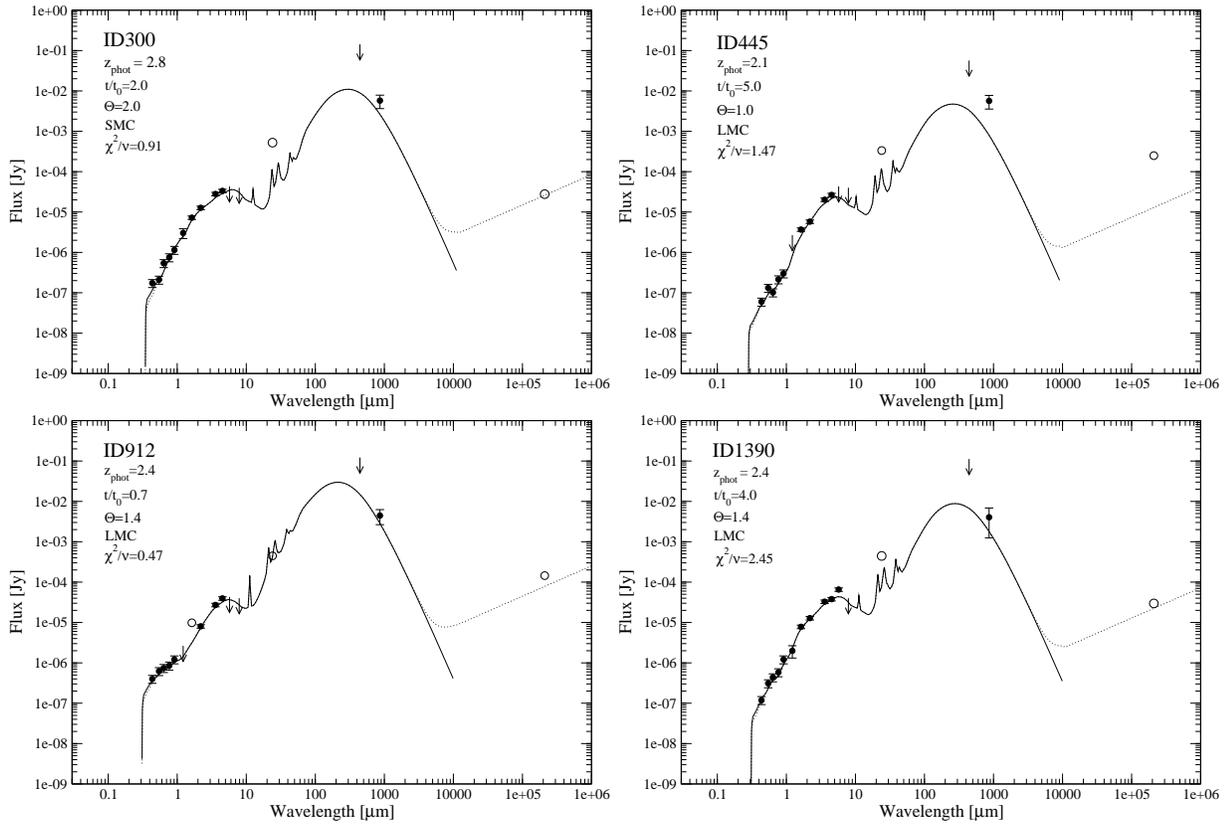

  \resizebox{8cm}{!}{\includegraphics{figure/sed_id300_swire.eps}}
  \resizebox{8cm}{!}{\includegraphics{figure/sed_id445_swire.eps}}
  \resizebox{8cm}{!}{\includegraphics{figure/sed_id912_swire.eps}}
  \resizebox{8cm}{!}{\includegraphics{figure/sed_id1390_swire.eps}}
 \caption{Solid and dotted lines indicate the best-fitting SBURT 
model without and with an extended radio component of the SED, respectively. The data 
points used for the SED fitting are shown as solid circles. Arrows 
indicate the 5\,$\sigma$ upper limits. See Section 6.2.1 for the definition 
of the fitting parameters.
}
 \label{sed}
\end{figure*}

\begin{figure*}
  \resizebox{7cm}{!}{\includegraphics{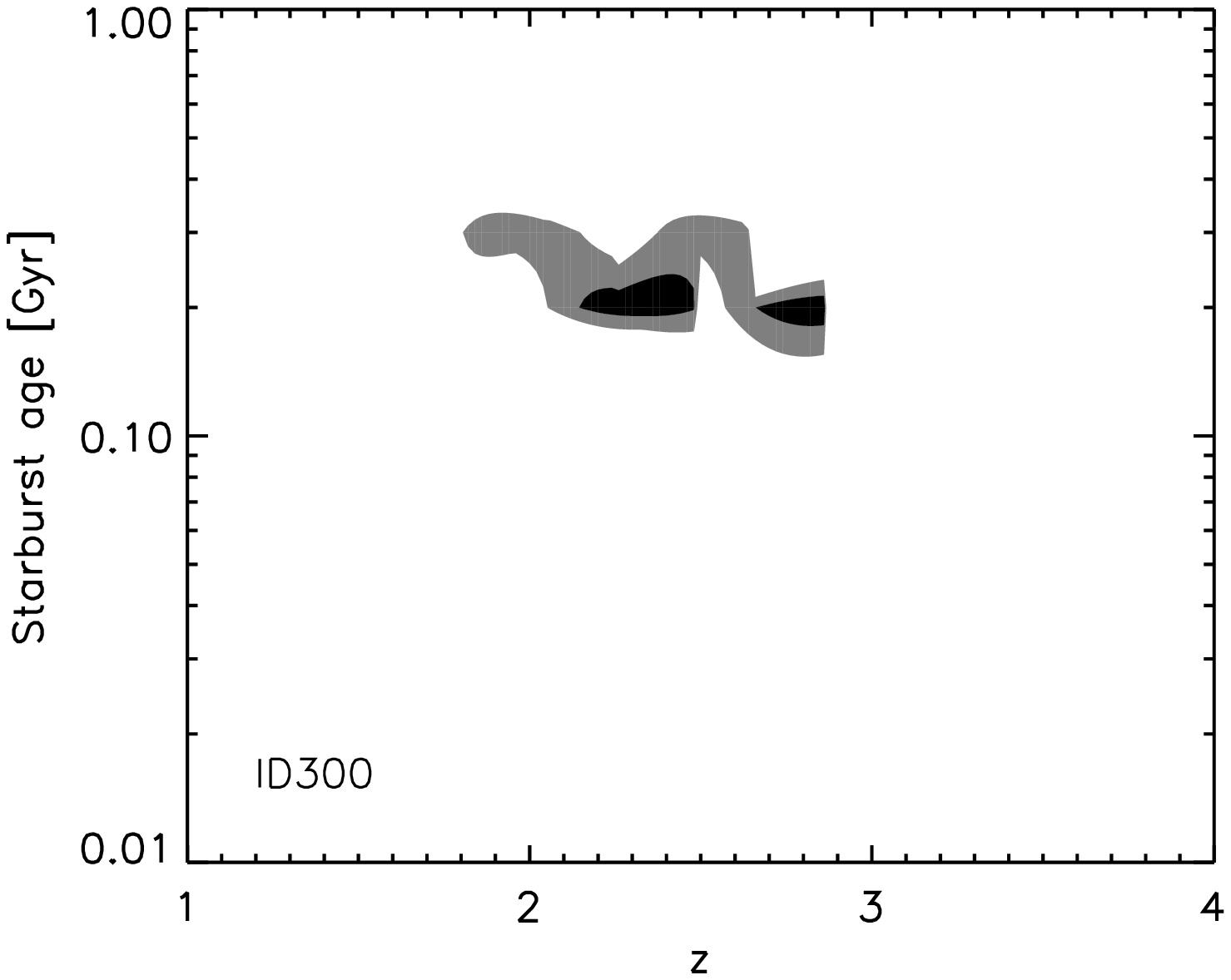}}
  \resizebox{7cm}{!}{\includegraphics{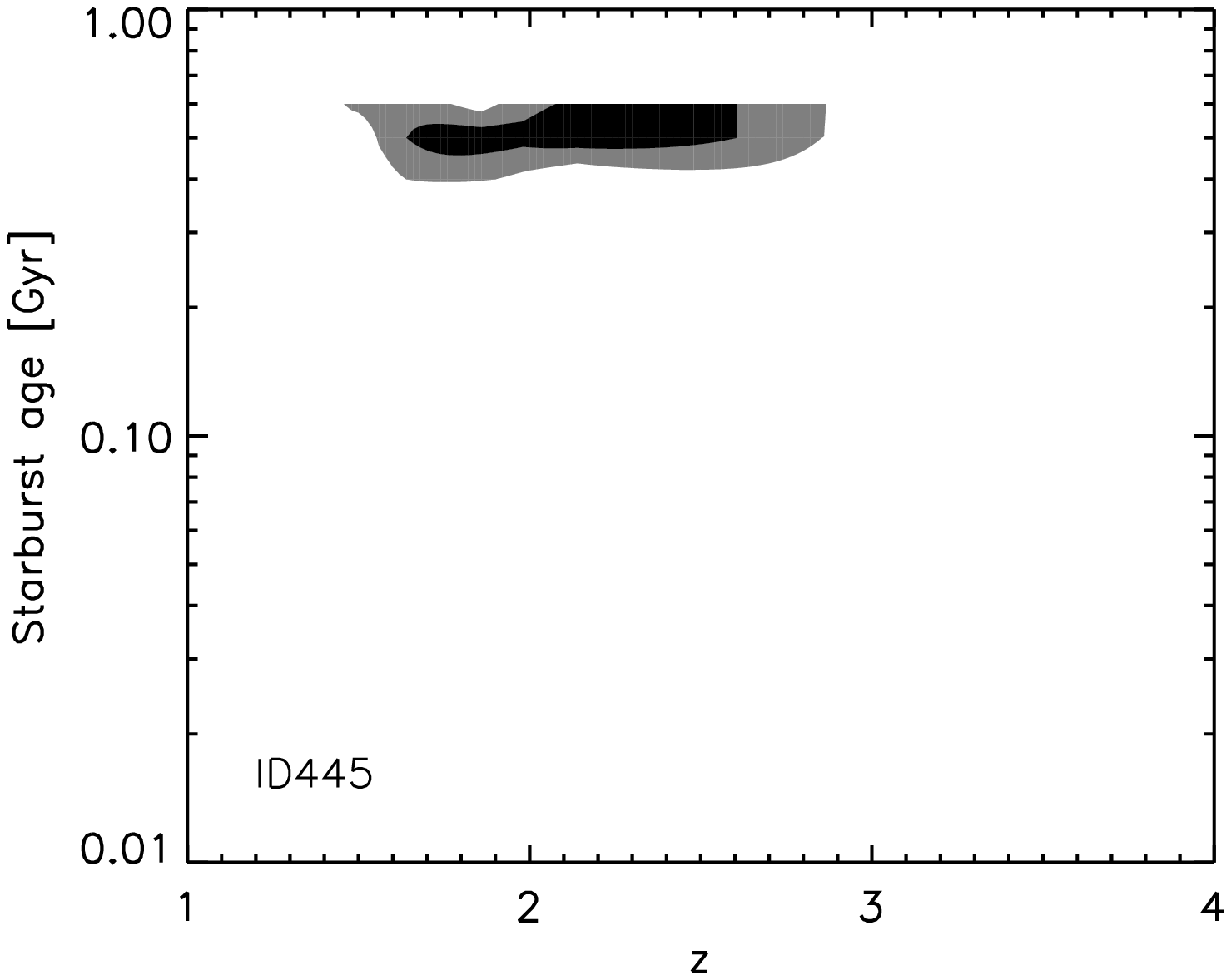}}
  \resizebox{7cm}{!}{\includegraphics{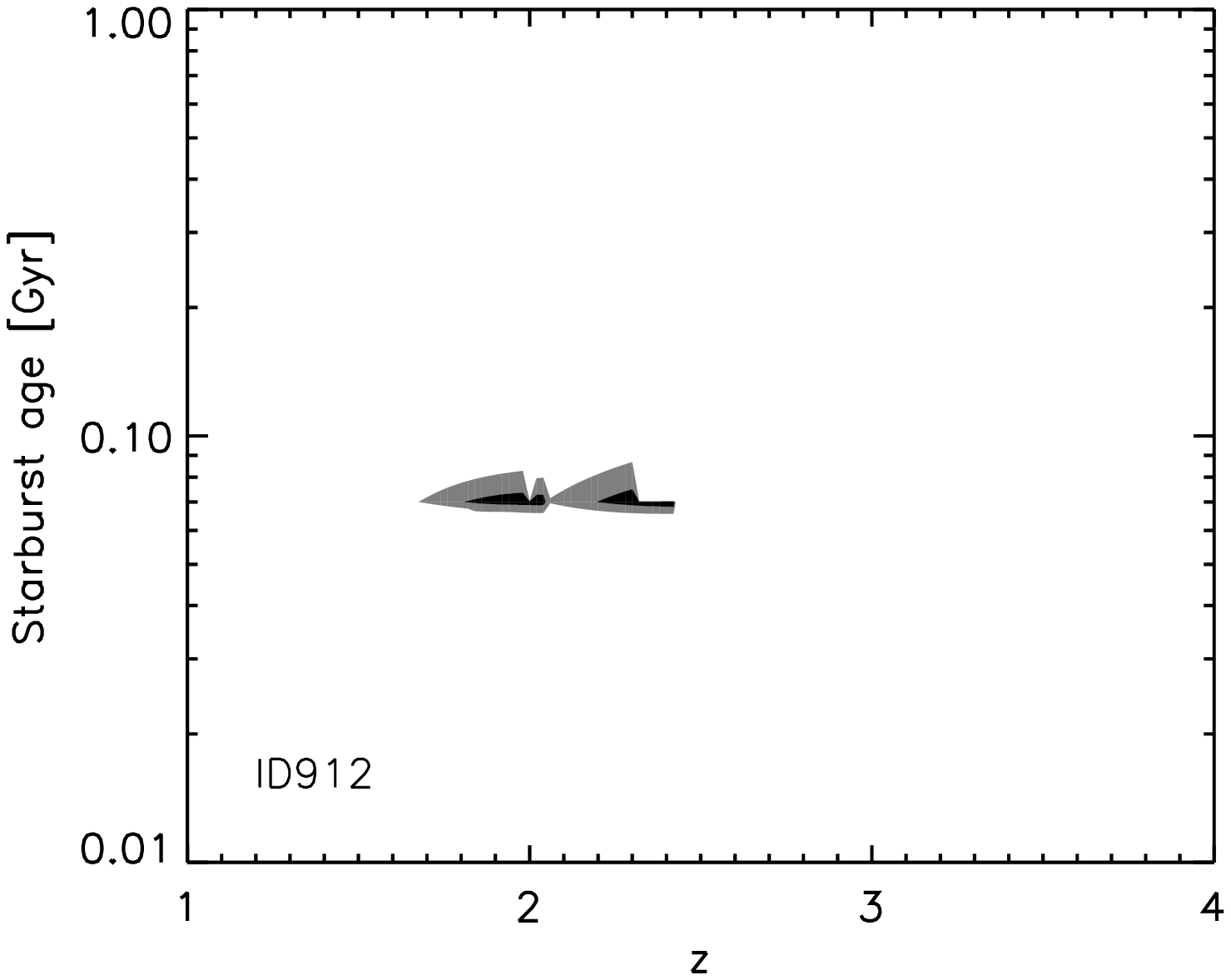}}
  \resizebox{7cm}{!}{\includegraphics{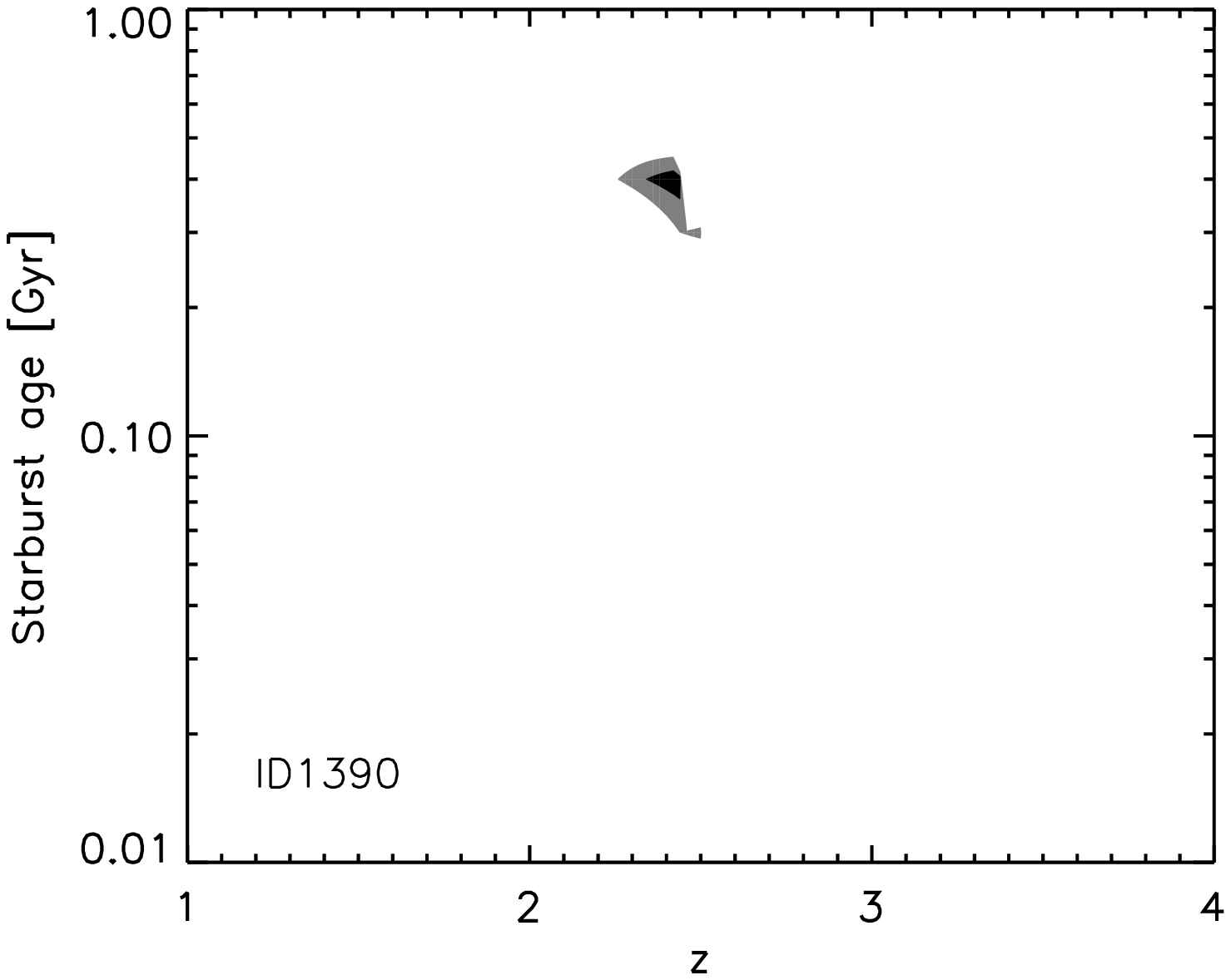}}
 \caption{
Contour maps of $\Delta \chi^2$ from the SED fitting in a plane 
of starburst age (with $t_0=0.1$\,Gyr)
and redshift. The contours are depicted 
at $\Delta \chi^2 =  2.71, 6.63$, having the probabilities of 90 and 99 
\% when projected on to each axis. We note that there are no local minima 
at $z<1$ and $z>4$. 
}
 \label{error}
\end{figure*}

\begin{figure*}
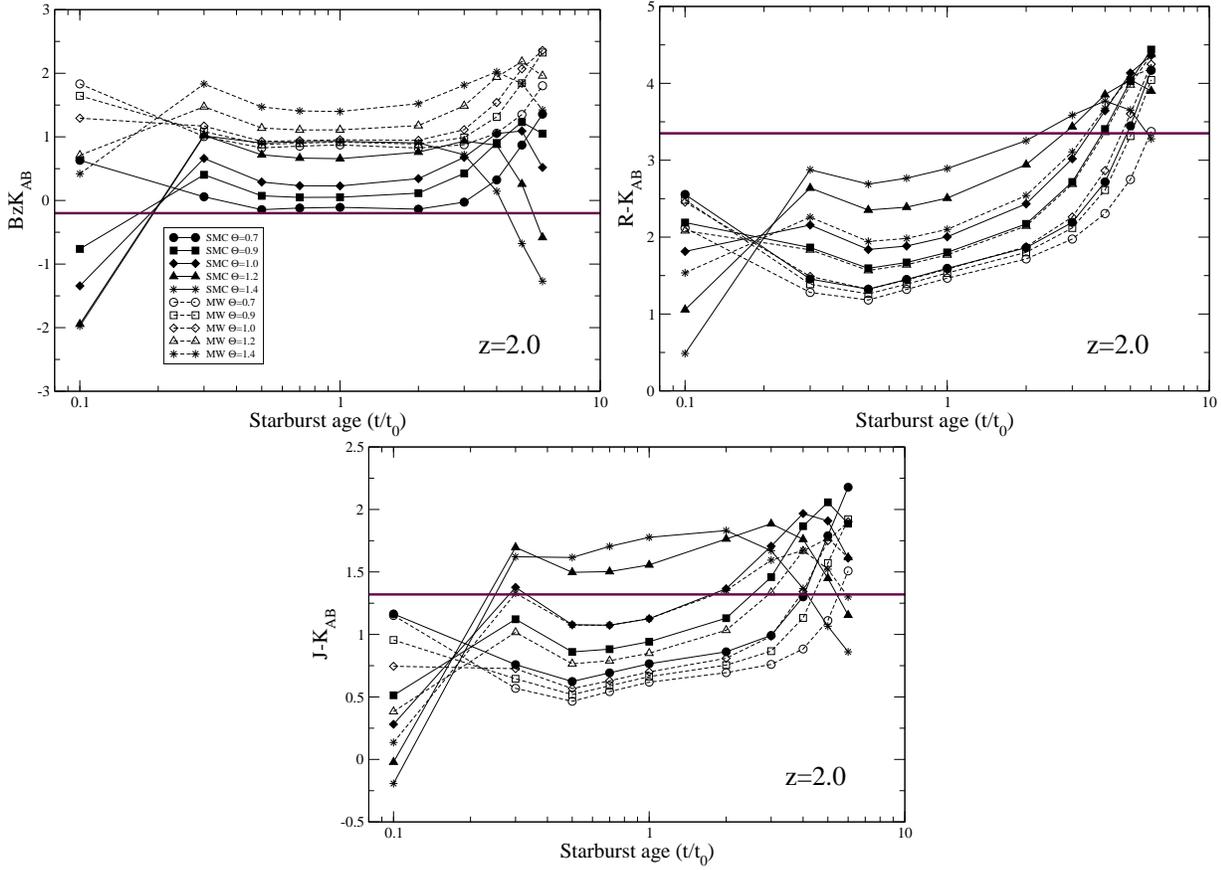

  \resizebox{8cm}{!}{\includegraphics{figure/colBzK_age_z2.0_sub.eps}}
  \resizebox{8cm}{!}{\includegraphics{figure/colRK_age_z2.0_sub.eps}}
  \resizebox{8cm}{!}{\includegraphics{figure/colJK_age_z2.0_sub.eps}}
 \caption{
$BzK$, $R-K$ and $J-K$ 
colours of the SBURT model at $z=2$ as a function of starburst age, which 
show the range of model parameters to satisfy $BzK$, DRG, and ERO selection 
criteria. Model parameters for a given line symbol are shown in the legend box. 
Horizontal lines indicate the colour boundary for $BzK$s, DRGs 
and EROs. 
}
 \label{bzk_col}
\end{figure*}

\begin{figure*}
  \resizebox{8cm}{!}{\includegraphics{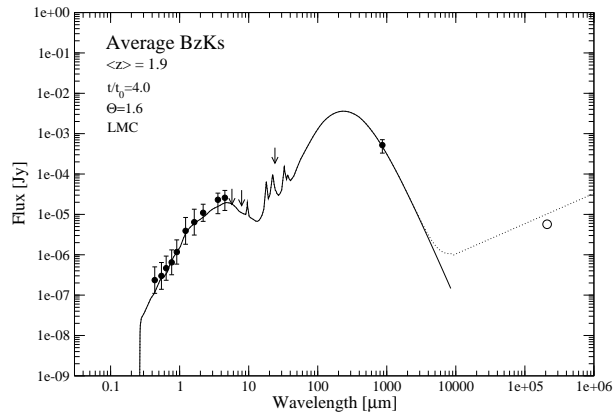}}
 \caption{Average SED of $BzK$s with $1.5 < (B-z') < 2.5$ and a representative 
SBURT model for the average SED. Symbols are the same as in Figure \ref{sed}. 
}
 \label{sed_ave}
\end{figure*}



\appendix 
\section{Estimate on the number of chance detections in our SCUBA maps}\label{appendix}
The large beam size of the JCMT  ($14.7''$ FHWM at 850\,$\mu$m) 
could lead to some chance detections of spurious 
$>$3$\,\sigma$ peaks at the position of NIR-selected galaxies. 
We statistically evaluate the expected number of $>$3$\,\sigma$ 
chance detections for a given class of galaxies as follows. 
The probability of randomly finding an object at $>$3$\,\sigma$ 
in the 850-$\mu$m map may be given by the ratio of the number of 
pixels with $>$3$\,\sigma$ to the total number of pixels in the 
region under consideration. 
For the SHADES/SIRIUS $K_\mathrm{s}$-band area, 
we found this probability as $p(>$3$\,\sigma)=0.45\times 10^{-2}$ 
for the SHADES B-map and $0.85\times 10^{-2}$ for the D-map\footnote{The 
difference in $p(>$3$\,\sigma)$ between the maps may be 
explained by the difference in the adopted pixel size.}.
If we randomly distribute $n$ objects in the 850-$\mu$m map, 
the number of objects spuriously detected with $>$3$\,\sigma$ would 
follow a Poisson distribution with the parameter of 
$\mu = n p(>$3$\,\sigma)$. For 307 NIRGs, 
we derive $\mu = 1.4$ and 2.6 for the B- and D-map respectively, 
giving the mean number of $>$3$\,\sigma$ detections by chance. 
This means that we must confirm whether submm-detected NIRGs 
are genuine submm emitters or not by other means. Here we require a 
detection at 24\,$\mu$m or radio to identify submm-bright NIRGs, 
along with the detection in the SCUBA map. For example, only 15 NIRGs are 
found to be detected at 24\,$\mu$m. This results in $\mu=0.067$ 
and 0.13 for 24\,$\mu$m-detected NIRGs in the B-map and D-map, respectively. 
Thus, we expect no chance detections in the SCUBA map if NIRGs are 
detected at 24\,$\mu$m.

\end{document}